\documentclass[10pt,a4paper]{article}
\usepackage{amsmath}
\usepackage{amsthm}
\numberwithin{equation}{section}
\usepackage{graphicx,lscape}
\title{\Large \bf Quantifying the quality of peer reviewers  through  Zipf's law}
\author{ Marcel Ausloos$^{1,2,3}$, Olgica Nedic$^{4}$, Agata Fronczak$^{5,\#}$,  Piotr Fronczak$^{5,\ddagger}$}

 \date{$^1$  School of Management, University of Leicester, University Road, Leicester  LE1 7RH, UK;\\$e$-$mail$ $address$: ma683@le.ac.uk \\\vskip0.5cm
 $^{2}$eHumanities
group\footnote{Associate Researcher}$\;$, \\Royal Netherlands
Academy of Arts and Sciences (NKVA), \\  Joan Muyskenweg 25, 1096 CJ
Amsterdam, The Netherlands \\ \vskip0.5cm $^3$GRAPES\footnote{Group
of Researchers for Applications of Physics in Economy and Sociology}$\;$,
  rue de la Belle Jardiniere 483, \\B-4031, Angleur, Belgium \\$e$-$mail$ $address$:
marcel.ausloos@ulg.ac.be
\\ \vskip0.5cm
$^4$ Institute for the Application of Nuclear Energy (INEP),
University of Belgrade, Banatska 31b, Belgrade-Zemun, Serbia    \\$e$-$mail$ $address$: olgica@inep.co.rs \vskip0.5cm $^5$  Faculty of Physics, Warsaw University of Technology,  \\Koszykowa 75, PL-00-662, Warsaw, Poland \\($\#$) $e$-$mail$ $address$: agatka@if.pw.edu.pl   \\($\ddagger)$   $e$-$mail$ $address$: fronczak@if.pw.edu.pl  }
\begin{document}
 \maketitle

\begin{abstract}
  This paper introduces a statistical and other  analysis of peer reviewers in order to approach their "quality" through some quantification measure, thereby leading to some quality metrics. Peer reviewer reports for  the Journal of the Serbian Chemical Society  are examined. The text of each report has first to be adapted to word counting software in order to avoid  jargon inducing confusion when searching for the word frequency: e.g. C must be distinguished, depending if it means Carbon or Celsius, etc. Thus, every report has to be carefully   "rewritten". Thereafter, the quantity, variety and distribution  of   words are examined in each report and  compared to the whole set.    Two separate months, according  when reports came in, are distinguished to observe any possible hidden spurious effects. Coherence is found.
 An empirical  distribution is  searched for through a Zipf-Pareto   rank-size  law. It is observed that peer review reports are  very far from usual texts in this respect.  Deviations from the usual  (first) Zipf's law are discussed.
A theoretical suggestion for the "best  (or worst) report" and by extension "good (or bad) reviewer",  within this context,  is provided from an entropy argument,  through  the  concept  of  "distance to average" behavior.  Another  entropy-based measure also allows  to measure the journal reviews (whence reviewers) for further comparison with other journals through their own reviewer reports.

\end{abstract}
     \vskip0.5cm
 %PACS: ???
 \vskip0.5cm
   
\section{ Introduction}\label{Introduction}
 
Peer review is posing  many problems. How do behave reviewers? What characterizes good or bad reviewers, good or bad reviews,  not only from  technical or scientific points of view, but also considering its linguistic, form and content, features.

We dare to claim that the work of reviewers is often not appreciated enough, but it deserves much more attention as it enables: (i) proper valorization of new results, (ii) recognition of new versus already published results, (iii) recognition of unreliable and even false data, (iv) recognition of plagiarism and
misconduct, (v) professional and public alert in the case of very good or very bad results, experimental design, ethical approach, etc.  

Despite a variety of criticisms  (Wager and Jefferson 2001), the importance of the peer review process in maintaining and improving the quality of submissions has been widely acknowledged. More than 80 percent of surveyed academics agreed that journal peer review greatly helps scientific communication  (PRC 2008). Nevertheless, there appears to be little agreement about how to measure its quality and effectiveness.

Typical methods proposed in the literature include: (i) surveys among the editors or among the authors whose work had been reviewed  (Justice 1998, McNutt 1990, van Rooyen et al. 1999), (ii) measuring agreement among reviewers  (Oxman 1991, Strayhorn 1993) or between reviewers and editors  (Callaham 1998), (iii) measuring the number of errors that a review detected  (Godlee et al.  1998) and finally (iv) measuring the speed of review  (Jadad et al. 1998, Feurer  et al. 1994, Neuhauser and Koran 1989). The other group of methods assess the quality of reviews indirectly, by analysing the manuscripts undergoing evaluation, e.g. (v) comparing the manuscript quality before and after peer review  (Goodman et al.  1994), or (vi) tracking the popularity of rejected (and published elsewhere) and accepted manuscripts  (Siler et al. 2015).

Among all feasible quantitative measures of reviews' (whence reviewers') quality, the most valuable are those based on objective criteria, e.g. on bibliometric indicators. On the other hand, measuring comprehensibility, soundness or informational content of a report can hardly be done without introducing a bias, due to individual opinions of the survey respondents (editors or authors) on how a good report should look like  (Bornmann 2011).  

In the present paper, we would like to start a discussion on a possible measure that would allow to quantitatively and objectively assess the quality of the linguistic and informational content of a report.

We will start from a rather trivial and well-known fact that language requires a diversity of words to convey a wide range of information. Regarding reviews, one can think that reviewers which use many different words are those who deal with the manuscript from more different aspects than those who use "less rich" vocabulary. Maybe these reviewers make more effort to offer authors clearer explanations or even suggest a direction for correction. Moreover, the authors of work being evaluated often want that the referees  use a larger, more precise, vocabulary to improve comprehension in view of removing ambiguities. On the other hand, long reports do not always  (or necessarily) provide a large  amount of information. They could just contain repetitive, rephrased statements.  That is why only the  length of a review is not a good measure of its informational content.

In this context, we would like to recall that Zipf  (Zipf 1949) formulated an algorithm (the so-called Zipf's law) that allows the evaluation and quantification of deviation between diversity and redundancy of different texts. Zipf's law is based on the {\it Principle of Least Effort}, which proposes that there must be a balance between unification and diversity in a language, such that the number of elements should neither be so highly
repetitive that the communication would be too simple, nor so heterogeneous that there would be too many possible combinations making communication unclear and convoluted. 

Therefore, since an efficient review should be the one that expresses an appropriate balance between diversity and redundancy (neither too diverse nor too repetitive), we have applied Zipf's law to estimate the quality of reviews with respect to their informational content. Other metrics could be used (Ausloos 2012a, Ausloos 2012b,  Darooneh  and Shariati 2014, Febres and Jaffe 2014, Rodriguez et al. 2014, Dubois 2014), but we restrict ourselves to the Zipf approach.

Zipf's law states that the frequency of a word in the text is inversely proportional to its rank in the frequency table  (Hill 2004). For example, the most frequent word is used twice as much as the second most frequent word and three times more often than the third most frequent word. Zipf's law is formally written  
\begin{equation}
P(r) \sim 1/r^{\alpha}
\end{equation}
where $P(r)$ is the frequency of occurrence of the $r$-th ranked item and the $\alpha$ parameter, which is usually close to $1$, is estimated from the slope of the resulting straight line that the word data follows on a log-log scatter plot. A steeper line, $\alpha>1$ (a more negative slope), represents a smaller, more repetitive vocabulary that may be too restrictive to efficiently convey information in the text. On the other hand, a flatter line, $\alpha<1$, represents a more diverse vocabulary. 

The deviations of the exponent $\alpha$ values have been reported in many different contexts, e.g. in different forms of schizophrenia  patients (Ferrer i Cancho 2006) and  in children language  (McCowan et  al.  2002), or in scientific texts  (Fairthorne 1969, Ausloos 2013, Miskiewicz 2013, Bougrine 2014). Although, only a few papers are quoted here in order to pin point a few research aspects in the field, but many more exist,  none seems to be devoted to the analysis of the aspect of peer-review intending to quantify the peer reviewer writings, in a simple way, at first, as this article proposes. Necessarily, conclusions are to be discussed, whence later on improved,  as any of open stage papers.

At the moment we cannot tell how can certain numbers obtained by us for specific reports be scientifically interpreted, but it seems that this part of the peer-review process can be subjected to quantification. Yet, another aspect of peer-review quantification may emerge, as peer-reviewers, becoming aware that their reports are being "evaluated" along quantitative lines, may become more "serious" and/or "professional" when reviewing. Finally, in the light of the general quantifying trend in science, perhaps one
day the quality of the peer-review activity may be evaluated and expressed by numbers.

The paper is organized as follows: 
  
In Section \ref{methodology}, the methodology  is presented: (i) data acquisition, containing  some information on   the data origin  (Section  \ref{dataacquisition}); (ii) demonstratoin that a necessary data refinement  for adaptation to available word counting softwares (Section  \ref{datarefinement}).  Thereafter,     a coherent  data analysis of the  various cases provided for  the illustration, with the assessment of some rank-size rule fits, is performed in Section  \ref{dataanalysis}.  It  will be observed and emphasized  that peer review reports are far from usual literary texts.
 
Section \ref{sec:results} is devoted to a subsequent analysis of  the investigation: within this data and following such results, an attempt is made to differentiate  reports from reviewers. This is also made, when possible,  i.e. in a few cases, by searching for similarities about reviewers having written reports  for different papers.

A thought on entropy consideration  to measure reports and reviewers through some so called "distance"   notion is presented in Sect. \ref{sec:entropy}, in order to suggest some useful  ("universal") metrics.

Section \ref{conclusions} allows us to conclude that each report definitely depends on the reviewer, but not  especially on the paper content.  We offer suggestions for further research lines.

\section{Methodology}\label{methodology}
\subsection{Data Acquisition}\label{dataacquisition}

At first, it is recognized that it is not easy to obtain raw data, even if anonymity is strictly enforced! However, one sub-editor of a section of the Journal of the Serbian Chemical Society (JSCS) has provided us with about 100 among the  latest reports arrived in the fall 2014, about papers submitted to the Biochemistry and Biotechnology section of the JSCS. In this data, the names of the referees have been anonymized and replaced by numbers, letters and symbols.

Two sets of reports have been examined: (i) for September and (ii) for October 2014.  For September, ten reports  $R_i$, with $i=1, \dots, 10$,  were selected: the last 10 which arrived. They correspond to   various reviewers  ($Q$) and topics, and to
six different papers ($P$): paper  $P_1$ was reviewed in $R_1$, $R_2$, and $R_3$; paper $P_2$ in $R_4$;  paper $P_3$ in $R_5$ and $R_6$; paper $P_4$ in $R_7$ and $R_8$; paper $P_5$ in $R_9$, % (me, rejection) 
   and paper $P_6$ in $R_{10}$. % (me, rejection).
%Papers  $P_1$ to $P_4$ were  accepted ({\bf not important,  except for allowing a reference to a forthcoming paper, but  over  some time interval.  Another point might be "When they were reviewed", in what year(s) }, but papers  $P_5$ and $P_6$ were rejected. 

Except for $P_5$ and $P_6$ having had one common reviewer (the fact that was not known at the beginning of the study), all others had different reviewers.  

Ten other reports, $R_i$, with $i=11, \dots, 20$,  were chosen for October 2014.  This  number was chosen to be equivalent to that of  the first set for statistical purposes.  The  October selection was made out of 100 reports. Due to this large number of reviews,   a word  scale effect, with both longer and shorter reports, could also be searched for. Of course,  longer  reports were more intriguing: we expected more reliable conclusions with "apparently  more  serious"  peer-reviewers, than shorter reports, usually mentioning a positive statement, like "paper to be accepted for  publication (due to whatever reason).

Moreover in order to test a  possible personal effect specific to reviewers, we added, from the October set,  2 more reviewers who performed reviews of 2 different papers. Those were selected on  the initial observation  that they were  of interest because  having  a very large number of words.
 %, before text adaptation as discussed below in Sect. \ref{datarefinement}. 
  However, the truly  two longest reviews, called   $R_{0}$ and $R_{21}$, were \underline{$not$} selected at this level.  Indeed, $R_{0}$ mainly contains a rather  long list of 31 references which the reviewer wanted the authors to include.  However, this long list does not bring much in terms of word statistics since it mainly contains titles of papers, thus with chemical compounds, and journal titles. The report also contains several remarks on spelling/grammar errors in the manuscript, thus often  misprints, leading to the appearance of  many single words, used once.   However, the "long" $R_{21}$ has next been retained, because it appeared to pertain to   the paper   $P_{21}$, whence could be   compared to  a shorter $R_{22}$ set of comments on this same paper,  $P_{21}$.%; it seems of interest  to compare two different reports on the same paper as done for a few case for the Sept. 2014 month.  

For broadening the discussion, three other reports are  considered: $R_{31}$, $R_{32}$, and   $R_{33}$.  Reports  $R_{31}$ and $R_{32}$ refer to different papers, $P_{31}$ and  $P_{32}$,  but have been reviewed by the same reviewer.  Furthermore,  a shorter $R_{33}$ has been  added to the selection because it was about the paper $P_{33}$, reviewed by  the same reviewer who wrote  $R_{11}$ for $P_{11}$. 

In summary, there are 10 cases for which  a single report corresponds to a single  paper; 4 cases in which a paper has been reviewed by a multiple set of reviewers; and 3 cases in which a submitted paper has  been reviewed  by different reviewers. Thus,   25 reviews ($R_j$) with 22 reviewers ($Q_k$) for 20 papers ($P_i$) are studied here.

We stress that authors and reviewers anonymity has been preserved throughout. We guarantee that, at first, there was no information used concerning the fact that a reviewer might have been the same for different papers. Only on the third stage of selection of reviews, when we added 5 papers,  as emphasized here above, had  this information been (necessarily) taken into account for the selection. 

Secondly, the outcome of the reports, i.e. whether the assessed papers were accepted, rejected or whether a revision was suggested, was unknown at the beginning of the study. Since, at first glance, one can guess that accepted papers correspond to short reports, it is tempting to judge by the report length only whether the reviewer's attitude is positive or negative. However, if the paper is nonsense or it has been already published by the same or other authors, then short negative reports are also probable. This fact also favors  a quantification analysis approach based on the \underline{distribution} of words, as the \underline{number} of words itself is not sufficient to evaluate reviewers. 

It is also worth to mention that, for JSCS, all reviewers are chosen according to their expertise, i.e. they are picked up from SCOPUS database as professionals (experts) in a topic of the submitted manuscript. In this respect, it is unlikely that short reports and small number of words are attributed to insufficient knowledge of chemistry by reviewers.

In summary, in the presented analysis neither the fate of the paper has been considered, nor the conclusive recommendation by the reviewer. In these respects, we can consider that our study was a "blind one", unbiased.

\subsection{Data Refinement}\label{datarefinement}

In order to count the number of words for distinguishing reports and  hopefully reviewer behaviors, some slight modification of the report had to be made   in order to adapt the  review containing technical jargon for available word counting softwares.  An important technical point has to be first mentioned: the algorithm  does not recognize  greek or cyrillic  letters, subscripts,  digits nor  indices,    e.g. $Tm$ is equivalent to $T_m$; the lowercase letters  are equivalent to capital letters; the   mathematical and grammatical symbols, like   "-" , "+",  "  ' ", "$\&$", or "/"  are replaced by blanks.  

Since the greek letters $\alpha$, $\beta$, and $\mu$ do represent some information  about a chemical compound,  (e.g., $\alpha$-sheet; $\beta$-helix), these letters must be considered as words, rather than as a letter. The same holds for $k$ which might be  for  kilo, but corresponds also  to a chemical element   Potassium,  but can also mean  Kelvin,  or  for $C$   which can refer to Carbon or Celsius or a "constant"; or $m$, an abbreviation for some unit (milli). Also a chemical unit $mU$ must be distinguished from the greek letter $\mu$
% (this occurs in $R_6$) 
 which could also mean $micro$ when some unit is mentioned.  All reports have  been adapted to take into account such considerations. Others can be briefly listed and justified:  for example,
\begin{itemize}
\item the greek letters $\alpha$, $\beta$,  $\mu$  have been replaced by $alpha$,  $beta$, $mu$;
\item $kJ$  (and similar  units) have been kept as   specific words, when there is no ambiguity; 
\item it has been checked whether $c$ (or $C$)  is Carbon or Celsius  or a "constant" or some specific symbol for a chemical species, and appropriately rewritten;
\item it has been checked whether $a$ (or $A$)  is an article or some specific symbol for a chemical species; 
\item and whether $K$ is Kelvin or Potassium, i.e. the $K$ for Kelvin has been kept, but $K$ for Potassium was replaced by "Potassium";
\item an identification was made  between Pb and Lead; to be distinguished from the verb "to lead";
\item  dash ($-$) presence was carefully checked, in order to replace the words by a single one when appropriate: e.g., $co-administration$ has been replaced by "coadministration", also in order to avoid a misinterpretation of  $co$ with Cobalt (Co);
\item   chemical compounds like Vitamin C (or Vitamin A) became  $VitaminC$ or  $VitaminA$; rGST-Mus became  RGSTMus; yet, $VPA-Induced$ are two words $VPA$ and $induced$, but $DT-diaphorase$ is only one word  $DTdiaphorase$;
\item   ionisation-mass spectrometry  has been kept as  a word set: ionisation mass spectrometry;
\item a $P$ for "Peak"  was distinguished from  "Phosphorus" or $p.$ for a page number;
\item   "$don't$" has been replaced by $donot$, in order to avoid confusion with  $T$, the temperature; idem for $didn't$ and $doesn't$ which read $didnot$ and $doesnot$, to keep  each of those  as the reviewer apparently wishes, i.e. a single word;
\item cyrillic letters have been arbitrarily simplified to read like in English alphabet - usually for author's names;
\item  when authors have complex names "El Alali" or "Gustin-Schwartz" they have been concatenated and  have been rewritten as ElAlali or GustinSchwartz;
\item initials of names have been attached to the names in order to avoid confusion with chemical compounds sometimes;    the names and initials of authors have been kept in the order given by the reviewer, since he/she wishes  so,  in some sense; for example  MDGardiner  differs from GardinerMD  (although it should be DMGardiner); this concatenation applies to references in the bibliography list as well;
\item the  words $et$ $al.$ have been forcefully replaced by $etal$, in order to  avoid confusion with Al for Aluminum;
\item  the numbers referring to Tables ($I$, $II$, ...) have been replaced by their arabic numeral, 1, 2, ... in order not to confuse the  Table number with the letter I;
\item  misprints have been kept,  including  those in comments by reviewers pointing misprints by (and to)  authors; 
\item the name of journals has been concatenated since they refer to a given unit;  otherwise "of"  and "Journal" would appear very often,  without any useful meaning for the word counting;
\item references to $http://...$ websites were deleted.
\end{itemize}

Other examples of  technical points can be quickly mentioned. Abbreviations have been kept as in the original reviewer report; e.g. $ref$ is  one word, different from $reference$. Both Edaravone and EDA are different words, though the same compound.   Special  confusions could be  seen to  occur   in other cases:  (i) a zeolite was called "zeolite A";  this  has been called $zeoliteA$  to avoid over-counting the "a" as an article (3 times in $R_{10}$);  (ii)   a $t-test$  is once mentioned (in $R_5$); it has been considered to be two words, this $t$ has been included as a $tee$  for not representing  $T$ as the temperature. 

Nevertheless, such possible  confusions or ambiguities  (and maybe others which have been overlooked) can be considered as  minute effects on the overall  analysis, discussion, and conclusions. Yet, this cumbersome time consuming task  insures  more confidence in the following data analysis.

 \begin{figure}
%\centering
\includegraphics[height=16.8cm,width=14.8cm] % [height=14cm,width=15cm]
 {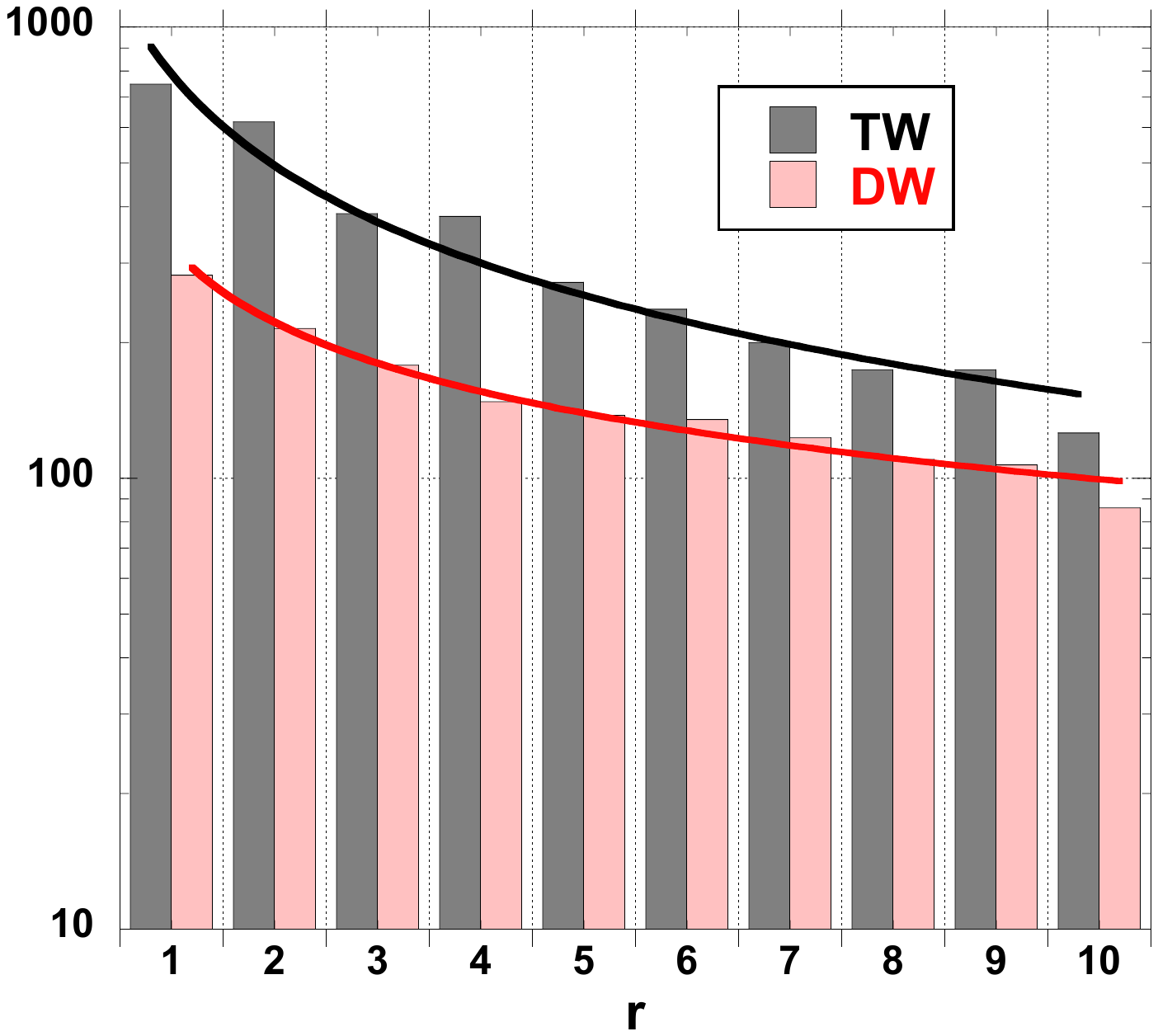}
 \caption{Rank-size relationship for the total number TW  of  words  and the number of different words DW used in the ten Sept. 2014  reports   each independently ranked by decreasing order of "importance", i.e. according to their TW or DW number;  fits are by a \underline{power law}  function; their corresponding  regression  coefficient is given.} \label{Plot10TWUWvsrpwlfits}
\end{figure}
 \begin{figure}
%\centering
 \includegraphics [height=16.8cm,width=14.8cm] % [height=14cm,width=15cm]
 {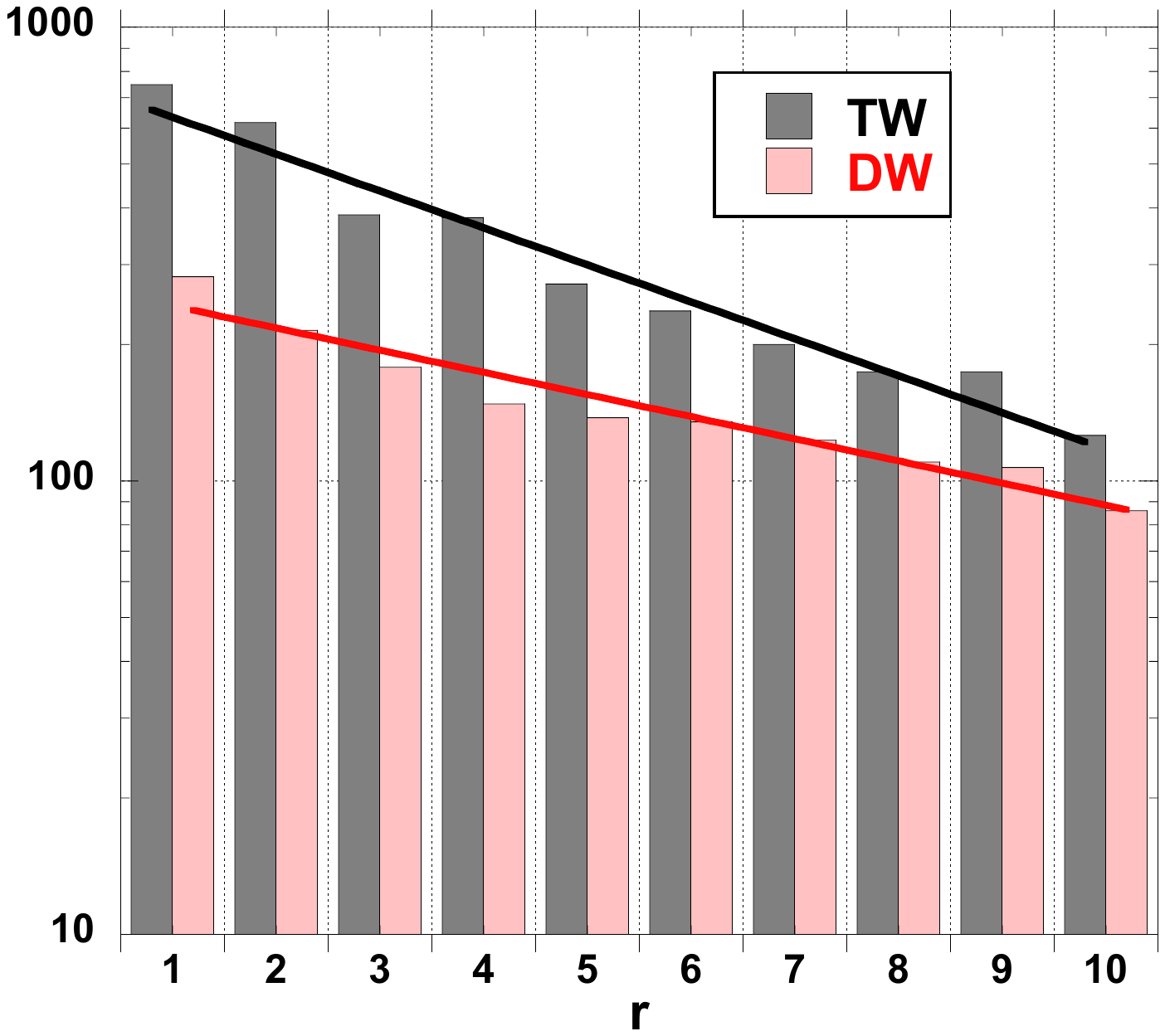} 
\caption{Rank-size relationship for the total number   TW  of  words and the number of different words DW used in the ten Sept. 2014  reports   each independently ranked by decreasing order of "importance", i.e. according to their TW or DW number;  fits are by   an \underline{exponential }function;  their corresponding  regression coefficient is given.} \label{Plot10TWUWvsrexpfits}
\end{figure}

% \begin{figure}
%\centering
%\includegraphics[height=12.8cm,width=10.8cm] % [height=14cm,width=15cm]
%{Plot6TWDWoctpow .pdf}
% \caption{Rank-size relationship for the total number TW  of  words  and the number of different words DW used in the ten  Oct. 2014  reports   each independently ranked by decreasing order of "importance";  fits are by a power law  function; their corresponding  regression  coefficient is given.} \label{Plot6TWDWoctpow}
%\end{figure}
% \begin{figure}
%\centering
% \includegraphics[height=12.8cm,width=10.8cm] % [height=14cm,width=15cm]
% {Plot6TWDWoctexp .pdf} \caption{Rank-size relationship for the total number   TW  of  words and the number of different words DW used in the ten Oct. 2014  reports   each independently ranked by decreasing order of "importance";  fits are by   an exponential function;  their corresponding  regression coefficient is given.} \label{Plot6TWDWoctexp}
%\end{figure}

       \begin{figure}
%\centering
\includegraphics[height=16.8cm,width=14.8cm] % [height=12cm,width=14cm]
{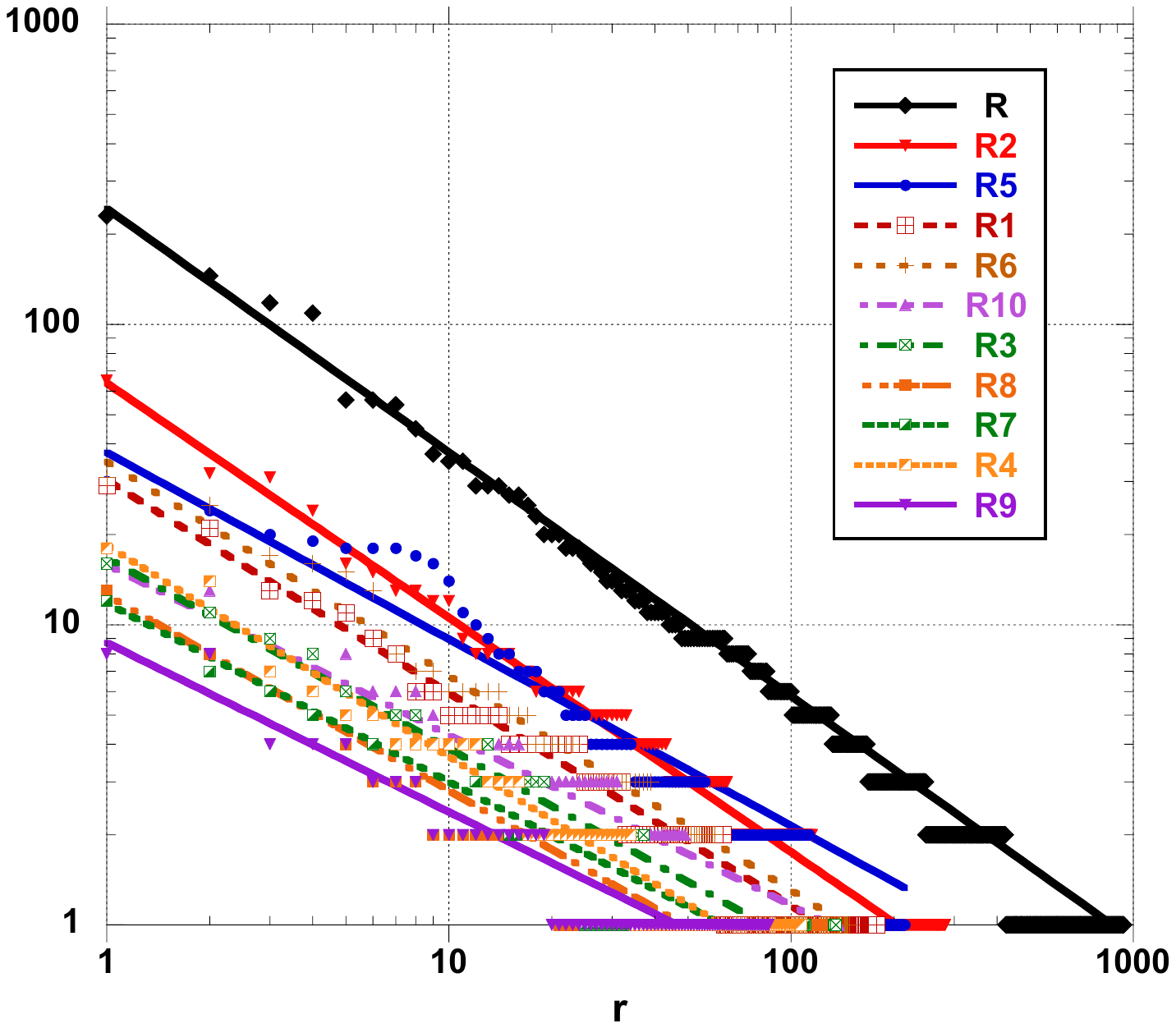}
 \caption   {Log-log  plot of the number  of different words used in the (10) Sept. reports $R_{1}$ to $R_{10}$, and the overall case $R_{1-10}$, as a function of their DW rank; fit parameter values are found in  Table  \ref{pwlfitparamSeptOct}.}
 \label{Plot25pwlfit11}
\end{figure}

       \begin{figure}
%\centering
\includegraphics[height=10.0cm,width=10.8cm] % [height=12cm,width=14cm]
{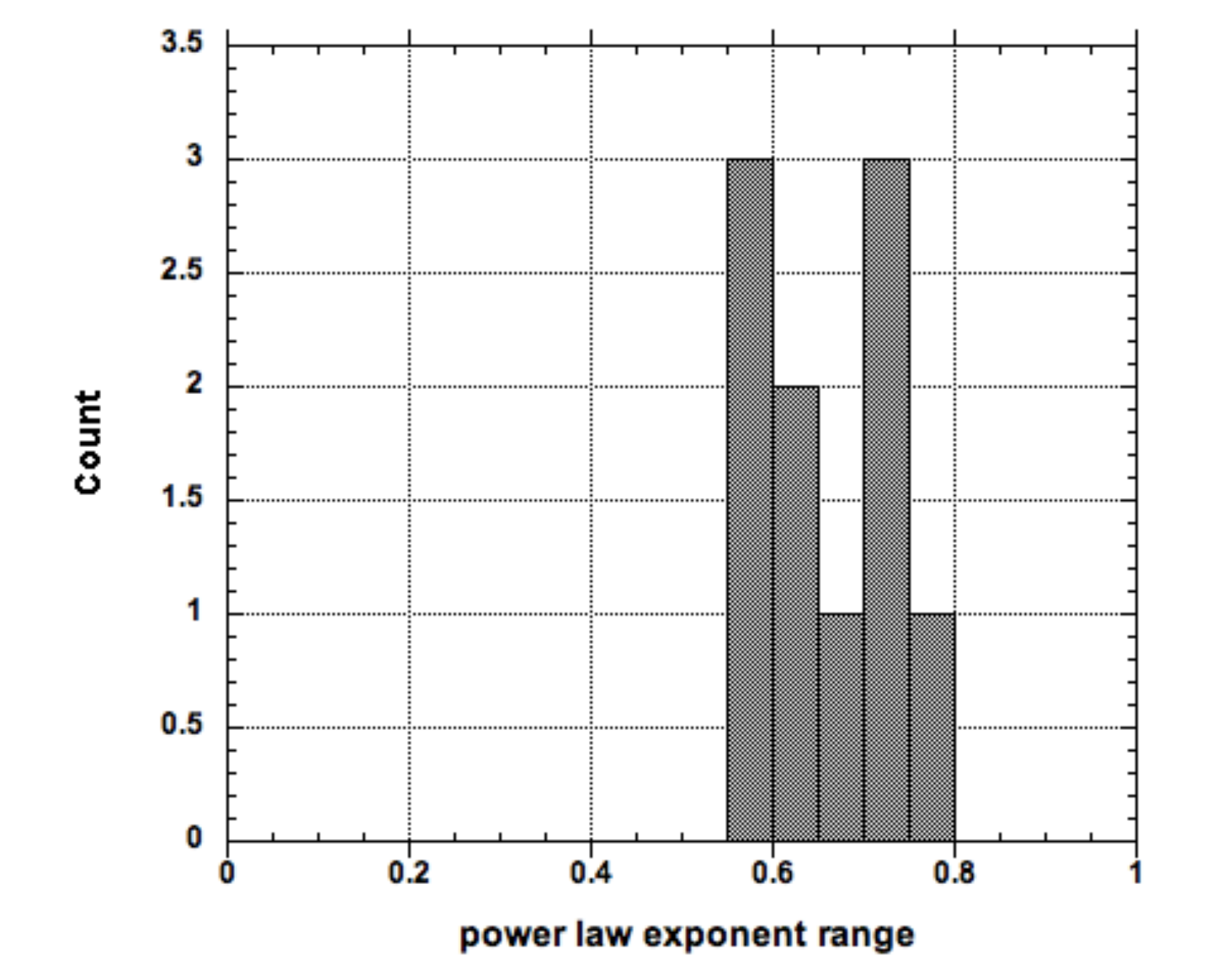}%{Plot6historaiRi.pdf}
 \caption   {Histogram of  the $\alpha$ exponent values for the rank-size relationship, Zipf's law,  Eq.(\ref{Zipfeq}), of the ten Sept. examined reports.}
 \label{Plot6historaiRi}
\end{figure}

    \begin{figure}
%\centering
\includegraphics[height=16.8cm,width=14.8cm] % [height=12cm,width=14cm]
{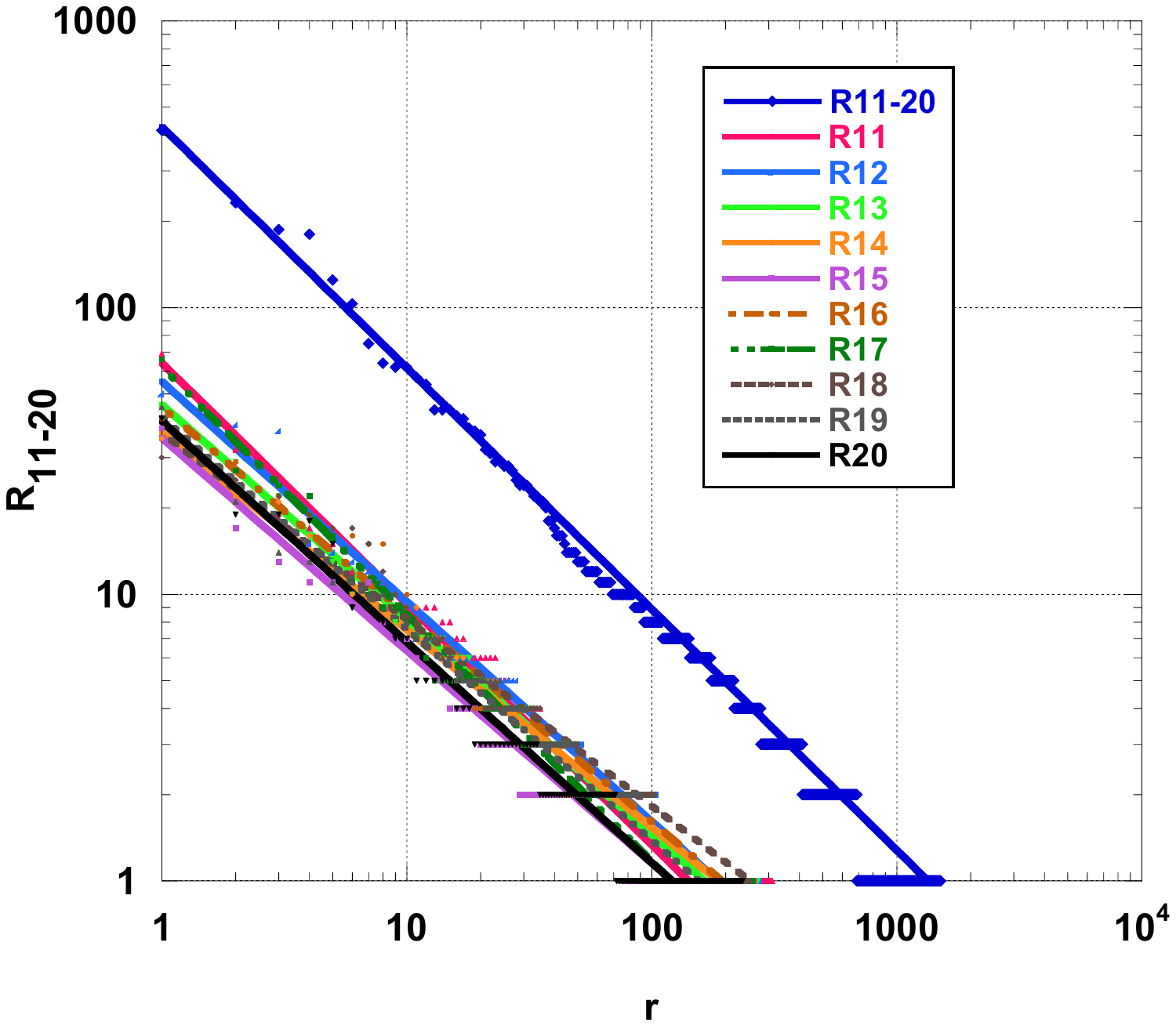}
 \caption   {Log-log  plot of the number  of different words used in  the (10) Oct. reports $R_{11}$ to $R_{20}$, and the overall  case  $R_{11-20}$, as a function of their DW rank; fit parameter values are found in  Table  \ref{pwlfitparamSeptOct}.}
 \label{Plot12R11R20fits}
\end{figure}

\begin{table}[htdp]
\begin{center}
\begin{tabular}{|c|c|c|c||c|c|c|c|c|c|c|c|}
\hline
% \multicolumn{2}{|c|}{ 2007 (PU) }&&  \multicolumn{2}{|c|}{ 2008 (RN) } 
$R_i$&	TW	&	DW	&	DW/TW &$R_i$&	TW	&	DW	&	DW/TW  \\ \hline 
$R_2$	&	746	&	280	&	0.37534	&	$R_{11}$	&	697	&	310	&	0.44476	\\
$R_5$	&	617	&	215	&	0.34846	&	$R_{12}$	&	663	&	277	&	0.41780	\\
$R_6$	&	386	&	148	&	0.38342	&	$R_{17}$	&	611	&	258	&	0.42226	\\
$R_1$	&	381	&	178	&	0.46719	&	$R_{18}$	&	595	&	251	&	0.42185	\\
$R_{10}$	&	269	&	139	&	0.51673	&	$R_{13}$	&	594	&	271	&	0.45623	\\
$R_3$	&	237	&	135	&	0.56962	&	$R_{16}$	&	594	&	261	&	0.43939	\\
$R_4$	&	200	&	107	&	0.53500	&	$R_{14}$	&	556	&	263	&	0.47302	\\
$R_8$	&	174	&	123	&	0.70690	&	$R_{19}$	&	538	&	232	&	0.43123	\\
$R_7$	&	174	&	110	&	0.63218	&	$R_{15}$	&	485	&	262	&	0.54021	\\
$R_9$	&	126	&	86	&	0.68254	&	$R_{20}$	&	475	&	235	&	0.49474\\\hline
$R_{01-10}$ 	 &	3310	&	937	&	0.28308	&	$R_{11-20}$	 &	5808	 &	1510	 &	0.25999	\\
  \hline			
\end{tabular}
\caption{The 20 reports ($R_i$) by reviewers ranked according to the  total number of words ($TW$) used for Sept. and Oct. 2014 respectively; the number of different words  $DW$  and the ratio DW/TW are    given, as well as these measures for the whole set of reports considered as a unique one} \label{TWDWR1to20}
\end{center} \end{table}

   \begin{table} \begin{center}
     \begin{tabular}{|c|c|c|c|c|c|c|c| c|c|c|c|c|c|c|c|c|c|c|c|c|c|  }
  \hline
 report	&	Max  	&	Mean& RMS	&	Std Dev	&	Var &	Std Err 	&	Skewn 	&	Kurt  \\ 	 \hline
$R_1$	& 	29 	&	2.14 & 3.76 &	3.11 & 9.68&	0.23  &	5.58	&	38.28	\\		
$R_2$	& 	65 	&	2.66 & 5.83 &	5.19 &27.02&	0.31	&	7.88	&	79.69	\\		
$R_3$	& 	16 	&	1.75 & 2.63	&	1.97 & 3.88&	0.16	&	4.40	&	23.65	\\		
$R_4$	& 	18 	&	1.86 & 2.95	&	2.29 & 5.26&	0.22	&	4.91	&	28.15	\\		
$R_5$	& 	30 	&	2.86 & 5.01	&	4.12 &16.99&	0.28	&	3.78	&	15.68	\\		
$R_6$	& 	31 	&	2.60 & 4.76	&	3.99 &15.96&	0.32	&	4.57	&	24.25	\\		
$R_7$	& 	12 	&	1.58 & 2.19	&	1.52 & 2.31&	0.14	&	4.00	&	20.28	\\		
$R_8$	& 	13 	&	1.41 & 2.04	&	1.47 & 2.17&	0.13	&	5.46	&	34.20	\\		
$R_9$	& 	8 	&	1.46 & 1.91	&	1.23 & 1.52&	0.13	&	3.86	&	16.47	\\		
$R_{10}$ &13 	&	1.93 & 2.76	&	1.97 & 3.91&	0.16	&	3.473	&	14.22	\\ \hline		
$R_{1-10}$&	230&3.53 &11.86 &	11.33 &	128.48&	0.37	&	13.08 &	214.47	\\		\hline
\end{tabular}
\caption{  Summary of statistical characteristics for the number of words ($N_c $) in the various  $R_i$ in September.}
\label{TablestattransposSept}
\end{center} \end{table}

   \begin{table} \begin{center}
     \begin{tabular}{|c|c|c|c|c|c|c|c| c|c|c|c|c|c|c|c|c|c|c|c|c|c|  }
  \hline
 report	&	Max  &	Mean& RMS	&	Std Dev	&	Var 		&	Std Err 	&	Skewn 	&	Kurt  \\ 	 \hline
$R_{11}$	&	69	&	2.24	&	5.28	&	4.79	&	22.95	&	0.27	&	9.97	&	125.31	\\
$R_{12}$	&	50	&	2.39	&	5.24	&	4.67	&	21.81	&	0.28	&	7.32	&	61.54	\\
$R_{13}$	&	45	&	2.19	&	4.42	&	3.84	&	14.80	&	0.23	&	7.33	&	66.88	\\
$R_{14}$	&	35	&	2.11	&	3.90	&	3.29	&	10.82	&	0.20	&	6.03	&	46.55	\\
$R_{15}$	&	38	&	1.85	&	3.49	&	2.97	&	8.85	&	0.18	&	8.13	&	85.65	\\
$R_{16}$	&	41	&	2.27	&	4.59	&	3.99	&	15.98	&	0.24  &	6.03	&	44.17	\\
$R_{17}$	&	66	&	2.36	&	5.50	&	4.97	&	24.79	&	0.31	&	9.30	&	106.17	\\
$R_{18}$	&	30	&	2.37	&	4.28	&	3.57	&	12.77	&	0.22	&	4.83	&	26.62	\\
$R_{19}$	&	45	&	2.31	&	4.38	&	3.73	&	13.92	&	0.24	&	7.53	&	75.43	\\
$R_{20}$	&	41	&	2.02	&	4.07	&	3.54	&	12.58	&	0.23	&	7.32	&	66.80	\\ \hline
$R_{11-20}$	 &	415	&	3.84	&	15.95	&	15.49	&	240.03	&	0.39	&	17.39	&	385.84	\\
	\hline
\end{tabular}
\caption{  Summary of statistical characteristics for the number of words ($N_c $) in the various  $R_i$ for October. }
\label{TablestattransposOct}
\end{center} \end{table}

\begin{table}[htdp]
\footnotesize
\begin{center}
\begin{tabular}{|c|c|c|c|c||c|c|c|c|c|}
\hline
                     &                      $S_1$                 &                  $\alpha$          &	$\chi^2$	& $R^2$	&  &                    $S_1$                 &                  $\alpha$          &	$\chi^2$	& $R^2$	         \\  \hline
$R_9	$&	8.72$\pm$0.27 	&	0.564$\pm$0.015	&	10.51	&	0.919	&	$R_{11}$	&	64.28$\pm$0.59	&	0.841$\pm$0.006	&	143	&	0.980	\\	
$R_4	$&	18.38$\pm$0.35	&	0.703$\pm$0.011	&	18.75	&	0.966	&	$R_{12}$	&	55.28$\pm$0.88	&	0.767$\pm$0.009	&	297	&	0.951	\\	
$R_7	$&	11.81$\pm$0.26	&	0.595$\pm$0.010	&	11.75	&	0.954	&	$R_{17}$	&	45.95$\pm$0.01	&	0.863$\pm$0.007	&	38.2	&	0.990	\\	
$R_8	$&	12.57$\pm$0.30	&	0.651$\pm$0.012	&	16.53	&	0.938	&	$R_{18}$	&	36.76$\pm$0.37	&	0.651$\pm$0.008	&	54.7	&	0.981	\\	
$R_3	$&	16.71$\pm$0.23	&	0.635$\pm$0.007	&	11.34	&	0.978	&	$R_{16}$	&	35.15$\pm$0.43	&	0.723$\pm$0.007	&	69.6	&	0.970	\\	
$R_{10}$	&	15.85$\pm$0.38	&	0.567$\pm$0.010	&	33.19	&	0.940	&	$R_{13}$	&	44.94$\pm$0.57	&	0.750$\pm$0.004	&	122	&	0.971	\\	
$R_6	$&	34.99$\pm$0.64	&	0.714$\pm$0.010	&	85.74	&	0.964	&	$R_{14}$	&	62.33$\pm$0.63	&	0.689$\pm$0.005	&	131 	&	0.979	\\	
$R_1	$&	30.27$\pm$0.31	&	0.708$\pm$0.005	&	24.86	&	0.986	&	$R_{19}$	&	36.63$\pm$0.71	&	0.737$\pm$0.006	&	195	&	0.939	\\	
$R_5	$&	37.48$\pm$1.01	&	0.622$\pm$0.012	&	357.9&	0.902	&	$R_{15}$	&	41.41$\pm$0.48	&	0.742$\pm$0.007	&	75.8	&	0.976	\\	
$R_2	$&	63.79$\pm$0.45	&	0.781$\pm$0.004	&	79.30	&	0.990	&	$R_{20}$	&	40.18$\pm$0.49 &	0.770$\pm$0.007	&	78.6	&	0.973	\\\hline	
&	243.31 $\pm$1.17 	&	0.812$\pm$0.003	&	1767	&	0.985	&		 &	428.0$\pm$1.3	 &	0.841$\pm$0.002	 &	3594 	 &	0.990	\\ \hline	
\end{tabular}
\caption{Power law fit parameter, Eq.(\ref{Zipfeq}),  for the 20 reports ($R_i$) by reviewers, ranked according to the  total number of words (TW with $size$ $S_i$) respectively for the Sept. and Oct. samples. The last line  corresponds  to the  concatenation of ten reports, becoming $R_{1-10}$ and $R_{11-20}$, respectively.} 
\label{pwlfitparamSeptOct}
\end{center} \end{table}

\section{Data Analysis}\label{dataanalysis}

  Table \ref{TWDWR1to20} contains  the counting of words in the  Sept. and Oct. 2014 reports ($R_i$) by reviewers ranked in decreasing order of  the  total number of words $TW$ so used; the number of different words, $DW$, is also given.  Observe the inversion of  $R_1$ and $R_6$, as well as  $R_4$, $R_7$, and $R_8$, between the two Sept. lists.  A similar shuffling is found for the Oct. lists, with   $R_{13}$,  $R_{16}$, and  $R_{14}$. Observe the position of  $R_{15}$ for which the $DW/TW$ ratio is the largest. Such a reviewer though $not$ writing a long report has a wide variety of words to express his/her view.
  
There is apparently no simple relationship between the $TW$ and $DW$ used by a reviewer.  
  There is a large variety of  report (or reviewer)  types from the   $TW$ and $DW$ point of view. 
It seems that  $TW$ and $DW$ depend specifically  on reviewers, on their vocabulary, and maybe on their willingness to spend some  time of peer-reviewing.  Notice from the Sept. data, that it appears that both very long and very short reports can be received  for the same paper, with  either very detailed or very modest explanations and comments, thereby hinting us to quantify reviewer behavior, through word count at first.
  
The  respective  rank-size relationships  for $TW$ and $DW$ is shown in Fig. \ref{Plot10TWUWvsrpwlfits} and Fig. \ref{Plot10TWUWvsrexpfits} for the Sept. cases,   searching  whether the empirical law is either a power law or an exponential. From  a regression coefficient $R^2$ values, it appears that $DW$ is better represented by a power law ($R^2\simeq0.986$), but $TW$ by an exponential ($R^2\simeq 0.959$).  The same is observed for the Oct. 2014 data, with  $R^2\simeq0.914$  and $R^2\simeq0.942$ for  $DW$ and $TW$, respectively.  The  relevant  graphs are not displayed, for conciseness. There is no apparently immediate explanation for such a different behavior.  It can be  surely observed  from the histogram of words  
 that the $TW$ and $DW$ distributions of words span different ranges. % as should be expected.
  
About  the results in Table \ref{TWDWR1to20}, one could  induce that a good report made by a good reviewer should contain many words, and many meaningful words. Thus  if one plots $TW$ vs. $DW$, one should conclude that $R_2$ is "the best", and $R_9$  "the worst".  However, one could argue that the ratio $DW/TW$ is more meaningful,  the larger the better, since only meaningful words should then be reported; whence $R_8$  becomes "the best", and $R_5$ "the worst". On the other hand, maybe one should be concerned with  some sort of two dimensional measure with the carefully chosen wages scaling the importance of both dimensions $DW$ and $TW$. In Section \ref{sec:results}, we  propose a measure exploiting Zipf's law which seems to be an adequate way to tackle this problem.
 
For completeness,  a summary of the statistical characteristics for the distribution of the
number  of words for each report $R_i$  is given in Table \ref{TablestattransposSept} and Table \ref{TablestattransposOct}.  Observe that the skewness and kurtosis are both always positive,  indicating  the existence of a large number of  rare terms.

\section{Reports and reviewers. A discussion}\label{sec:results}
First, let it be re-mentioned  that Zipf  (Zipf 1949) observed that a large number of size distributions, $S_r$ can be approximated by a simple {\it scaling (power) law} $S_r = S_1/r $, where $r$ is the ranking (integer) parameter, with  $S_{r } \ge S_{r+1}$, (and obviously  $r<r+1$). A more flexible equation, with two parameters,  reading
\begin{equation}\label{Zipfeq}
S_r = \cfrac{S_1}{r^{\alpha}},
\end{equation}
is called the rank-size scaling law and has been often applied to
many "sizes" of "things"  (Hill 2004, Cristelli  et al.  2012). The particular (Zipf) case ${\alpha}=1$ is thought to represent a desirable situation, in which {\it forces of concentration balance those of decentralization}, what in our study means an appropriate balance between diversity and redundancy in reviewers reports. Such a case,  called the rank-size rule, has been frequently identified and
sufficiently discussed  elsewhere  (Lin 2010, McKean et al. 2009, Wieder 2009, Wolfe 2009, Wolfe 2010) to allow us to base much of the present investigation on such a simple law. 

In this context, let us consider the 10 September reports, the 10 October reports, and the reports specially added to the investigation, successively.

\subsection{September reports}\label{September reports}
Thus, let us display  the number  of different words used in each report  $R_i$, from  $R_{1}$ to $R_{10}$, and in the overall  set  $R_{1-10}$,    as a function of their  specific rank in each case, on a   log-log  plot,  Fig. \ref{Plot25pwlfit11}. The fit parameter values  for the power law are found in  Table \ref{pwlfitparamSeptOct}, with their standard  error bars.

Observe that 80 words are very frequent; occurring more than 10 times; thus  likely in all reports. The high rank words are likely specific to each report pertaining    (most likely) to a few  (most likely, bis) different, papers.

The $\alpha$ values  of the $R_i$ rank from $0.56 $ till $0.78$, with a mean  $\mu\sim 0.654$, itself slightly above the median ($\sim 0.643$); see Fig. \ref{Plot6historaiRi}. It should be obvious from such values that report by reviewers are far from classical texts; in these, the $\alpha$ exponent is usually close to 1 indeed.

Several other points are remarkable:
\begin{itemize}
\item $R_5$ is (very) anomalous;  it has a weak $R^2\simeq 0.90$, due to a marked shoulder near $r\sim 6$,  see Fig. \ref{Plot25pwlfit11}, but its $\alpha \sim 0.62$ is close to the average;
\item the correlation coefficient for $R_9$ is far from being considered large, but this  report contains very few words;
\item the largest $\alpha$ corresponds to the "all Sept. reports" case,   $R_{1-10}$; it is known that the highest $\alpha$ of a sum of power laws  is dragged by the highest $\alpha$ of the set, here $R_2$;
\item although the statistics is only based on ten reports, one can  recognize a two peak distribution of $\alpha$ values: one below 0.6, the other at 0.7 (Fig.\ref{Plot6historaiRi}).
\item quite interestingly, it appears that reports $R_9$ and $R_{10}$ are characterized   $quasi$ by  the same exponent $\sim 0.56$. Recall that this observation led us to inquire from the editor about whether they were due to the $same$ reviewer!  It is! By the way, the papers were rejected.\end{itemize}

\subsection{October reports}\label{Octoberreports}
 
 In Fig.\ref {Plot12R11R20fits}, a  log-log  plot of the number  of different words used in reports $R_{11}$ to $R_{20}$, together with the overall number, $R_{11-20}$, as a function of their rank is shown. The fit parameter values  for the power law are found in  Table  \ref{pwlfitparamSeptOct}. The overall set of reports, $R_{11-20}$, is markedly well represented by a power law with a very high $R^2$. 

At first sight, it appears that the reports can be positioned  in two groups, basically according to the distribution of words around the straight line fit on a log-log plot. Also,  $R_{13}$, $R_{19}$, and  $R_{20}$   weakly deviate from the hyperbolic rank-size relationship. $R_{19}$ and  $R_{20}$  have "not many" words as the other reports;   $R_{13}$  is an intermediary case in that sense. In these three cases, $\alpha \sim0.75$. The above also indicates that the statistics is better when the text is longer. 

However in several reports the word frequency distributions only weakly agree with a power law fit. In particular, two kinds of effects can be observed: a King with Vice-Roy effect (when the low rank data are much above the fitted line) and a Queen with harem effect (when the rare words, often referring to misprints, are too abundant)  (Laherrere and  Sornette 1998, Ausloos 2013. In that cases, a Zipf-Mandelbrot fit  (Fairthorne 1969), i.e. a $3$-parameter natural generalization of a $2$-parameter Zipf's law, would be more appropriate.  Its interest falls outside the framework of our paper and is not here examined further.  
  The King-ViceRoy cases are $R_{11}$, $R_{15}$, and $R_{17}$; note that   $R_{11}$ and  $R_{17}$,   have the largest $\alpha$ exponent  ($ \sim  0.8$) and a large number of words. The Queen-Harem effect is more marked in $R_{12}$, $R_{14}$, $R_{16}$, $R_{18}$; notice that this is the set of  reports having a lesser amount of words than  the previous reviews.   This exponent $\alpha\sim 0.7$ is in the lower part of the range interval. 

These effects (deviations from Zipf's mere power law)  are much tied to the behavior at low rank where the usual  (most common) English words appear; the highest ranks pertaining to rare words are often mentions of misprints.  Nevertheless, their presence indicates that the reviewer has seriously read the submitted paper.
This  confirms the interest of a word counting  analysis  beside the Zipf's line of approach.
 
 \subsection{Comparing reports by the same reviewer on different papers}\label{sub43}
Comparing reports by  the same reviewer on different papers  is not an easy matter, because it is a rather rare event. Nevertheless, 3 cases occur in our data set. Thus, a  comparison of the  Zipf power law exponent for three reviewers ($Q_1\equiv$ N, $Q_2\equiv$ E, and $Q_3\equiv$ C)   each having reviewed two (different) papers: $R_{31}$ and $R_{32}$,      $R_{11}$ and $R_{33}$,    and  $R_{21}$ and $R_{22}$,    
respectively, is made on Fig. \ref{Plot3revNEC}.  The  $\alpha$   exponents  are  remarkably different for the three reviewers,   but  are  very close to each other, whatever the report.  Thus, a marked identification can be apparently made of the reviewer according to his/her $\alpha$ exponent. 

  \subsection{Comparing reports on the same paper by different reviewers}\label{sub44}
\begin{figure}
%\centering
\includegraphics[height=12.8cm,width=14.8cm]{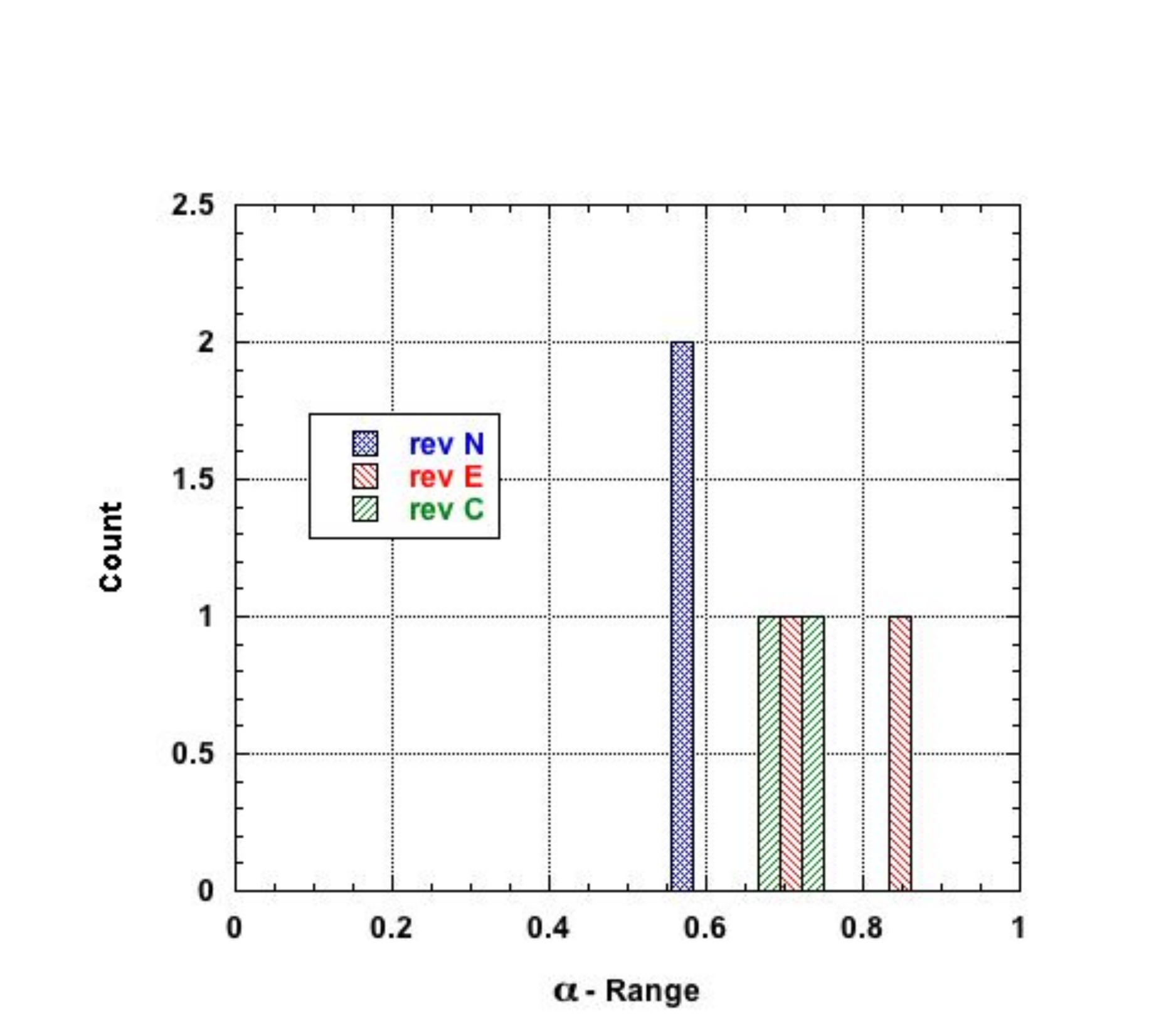}  % [height=14cm,width=15cm]
 \caption{Comparing Zipf power law exponent for three reviewers (for different  papers) indicating some coherence of a reviewer in his/her reports} \label{Plot3revNEC}
\end{figure}

\begin{figure}
%\centering
\includegraphics[height=12.8cm,width=14.8cm]{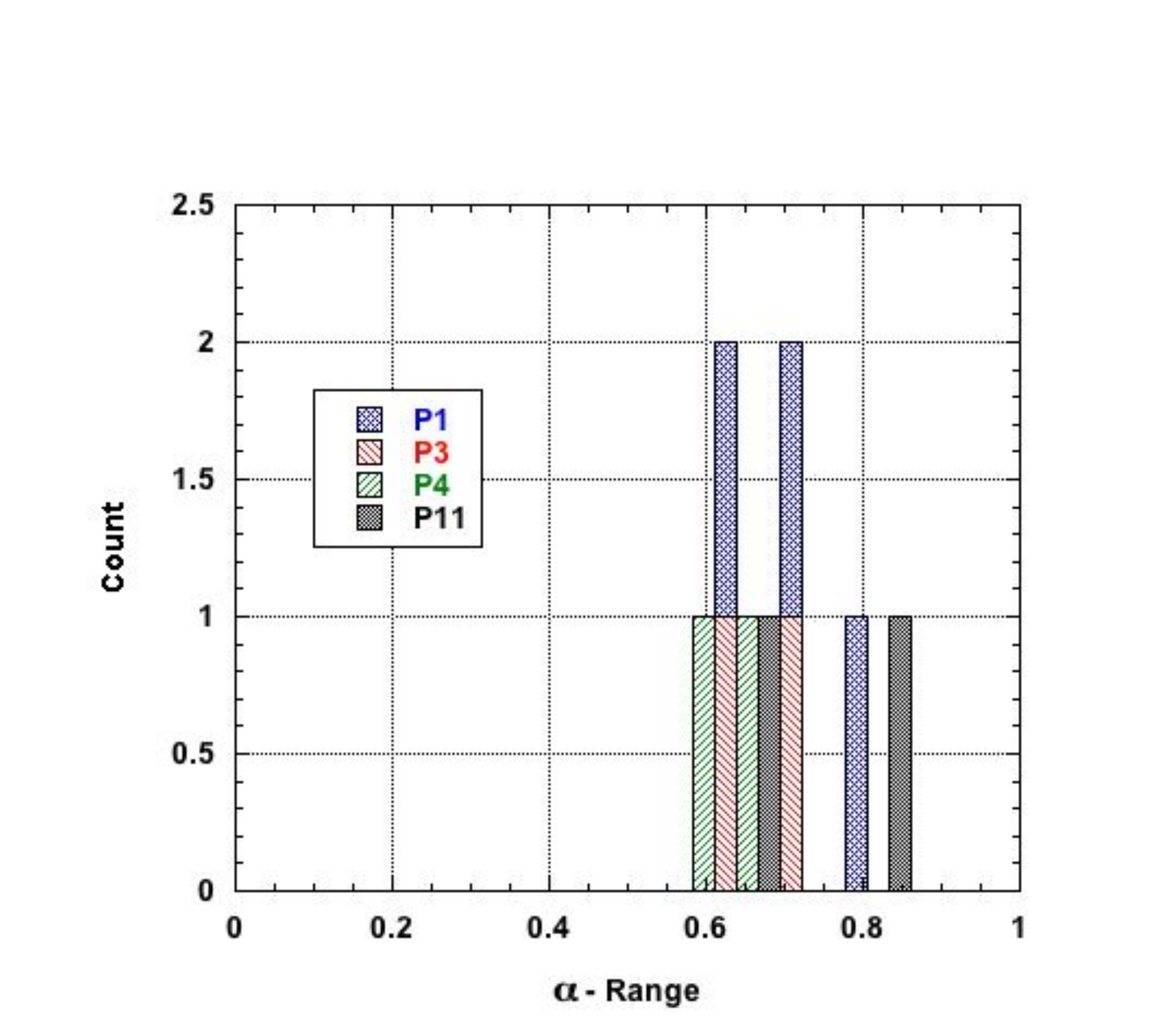}% [height=14cm,width=15cm]
 \caption{Comparing the Zipf power law exponent of different  reviewers for  the same  (as indicated) papers, hinting to consider a  weak influence of the paper  on  the reviewer characterization.} \label{Plot3revP1P3P4P11}
\end{figure}

Moreover,  for completeness, we can compare reports on the same paper by different reviewers. The Zipf power law exponents  for  different  reviewers for  the 4 indicated papers ($P_1$  through $R_{1}$, $R_{2}$, and     $R_{3}$; $P_3$ through   $R_{5}$ and $R_{6}$;  $P_4$ through $R_{7}$ and     $R_{8}$;  $P_{11}$ through $R_{11}$ and     $R_{33}$)  are shown on Fig. \ref{Plot3revP1P3P4P11}.   In contrast to Sect.\ref{sub43} and Fig. \ref{Plot3revNEC} a distinction of reviewers can hardly be made in this case, due to the grouping of $\alpha$ values on a short range interval..

\section{Entropy connection}\label{sec:entropy}
 
%{\it chen: the rank-size pattern comes from the process of entropy maximization, which is very important in urban modeling (Wilson, 1968; Wilson, 1971).  }

Recall that the exponent $\alpha$ is considered to be a  characteristic measure of the text content (and reviewer, by extension). %{\bf A new histogram of the distribution of $\alpha$ exponents, different from the one presented in Lisbon, is given on 
Its distribution for the 20 presently examined  reports %on Fig. \ref{Plot13histoalphaseptoct}                      
   emphasizes a peak at 0.75 or so,  but also the possibility of a multi modal structure !  % {\bf (THIS FIGURE COULD BE REMOVED)} 

The distribution of $\alpha$ exponents is  best studied through a normalization condition which allows to define the "probability" for finding a certain "$\alpha$-state", i.e.  a rank-size occurrence.
 Going further, one can imagine to have  access to  the probability of  a certain type of report ($R_j$) by a reviewer ($Q_k$),   at a certain rank or $\alpha$ value, through
 \begin{equation} \label{pr}
p(\alpha_j)  \sim  \frac{\alpha_j }{\sum_{j=1}^{j_M} \alpha_j } ,
\end{equation}
 where $j_M$  is the number of reports. 
 
 In presence of a large number of data points, a formula like 
 \begin{equation} \label{prpiotr}
p(\alpha_j) \sim \frac{N(\alpha_j)}{\sum_{i=1}^{j_M} N(\alpha_i)}
\end{equation}
 where $ N(\alpha_i) $ can be considered  as the density of $  \alpha_i  $ in some $i$ interval, can be equally interesting.

 Consider the 20 reports recorded in Table  \ref{pwlfitparamSeptOct}.
 %Since  $\sum_{i=1}^{i_M} \alpha_j $=14.074 (= 6.540 + 7.534),   
 Each $p(\alpha_j)$ can be easily obtained from Eq.(\ref{pr}). 
Thereafter,  one can obtain something which looks like   a contribution to a Shannon information entropy  (Shannon 1948, 1951)     for a given report $j$, 
  \begin{equation} \label{Hi}
  H_j\equiv   -p(\alpha_j)\; ln (p(\alpha_j)),
  \end{equation}  %The total entropy is $H= \Sigma_j\;   H_j \simeq 2.989$.   
  
   \subsection{Report ranking relative distances}\label{Reportrankinghat}
   To estimate the validity of an empirical  distribution, it is  practical  to  compare each ($H_j$) measure 
  to their related maximum disorder number,  i.e.   $ln (N_j)$, where $N_j$ is the number of different words  (DW, or  number of data points)  found in  the $R_j$ report (see values in  Table \ref{TWDWR1to20}). This technique allows     to "measure"  some information content of a report  $R_j$ (by a reviewer $Q_k$).% is using (and distributing in his/her report); 
%for $N_i =400$,  $ln(400)\sim 6$. 

Thereafter,  we define the relative "distance"   (of a report $j$)  to the maximum entropy (full disorder) as 
 \begin{equation} \label{d}d_j = 1-\frac {H_j }{ln(N_j)} \end{equation}%$\sim 1-0.24/6= 0.96$, 

In order to rank the reports "in a thermodynamic or information content" sense, with respect to an average report, one can define the  average  report entropy for a set of  $j_M$ reports as
  \begin{equation} \label{H}
 \hat{ H}\equiv \frac{1}{j_M} \;\sum_{j=1}^{i_M}  H_j= 
\; -  \frac{1}{j_M} \sum_{j=1}^{j_M}  p(\alpha_j)\; ln (p(\alpha_j)).
  \end{equation} 
%i.e., $\hat{H}  \simeq   0.15$. 
%i.e., $\hat{H}  \simeq 0.256 $. 
Thereafter, the "distances" with respect to the \underline{average report}  can be defined as  the distance between the  two entropies
    \begin{equation} \label{daverage}\hat{d_j}= -\frac {H_j }{ln(N_j)}+\frac {\hat{H} }{ln(N/j_M)}\end{equation}
 where $N$ is the total number of different words  found in the set of $j_M$ reports. %  (N.B. 1996 $<$ 937+1510). (N=1996) for 20 reports.
 
 %(DW = 937, here; $ln(937)\sim 6.843$), leading to  a set of values,  $ \hat{d_i}\simeq 0.016$,   since $ln(93.7) \simeq 4.54$. 

\begin{figure}
%\centering
\includegraphics[height=14.8cm,width=14.8cm]{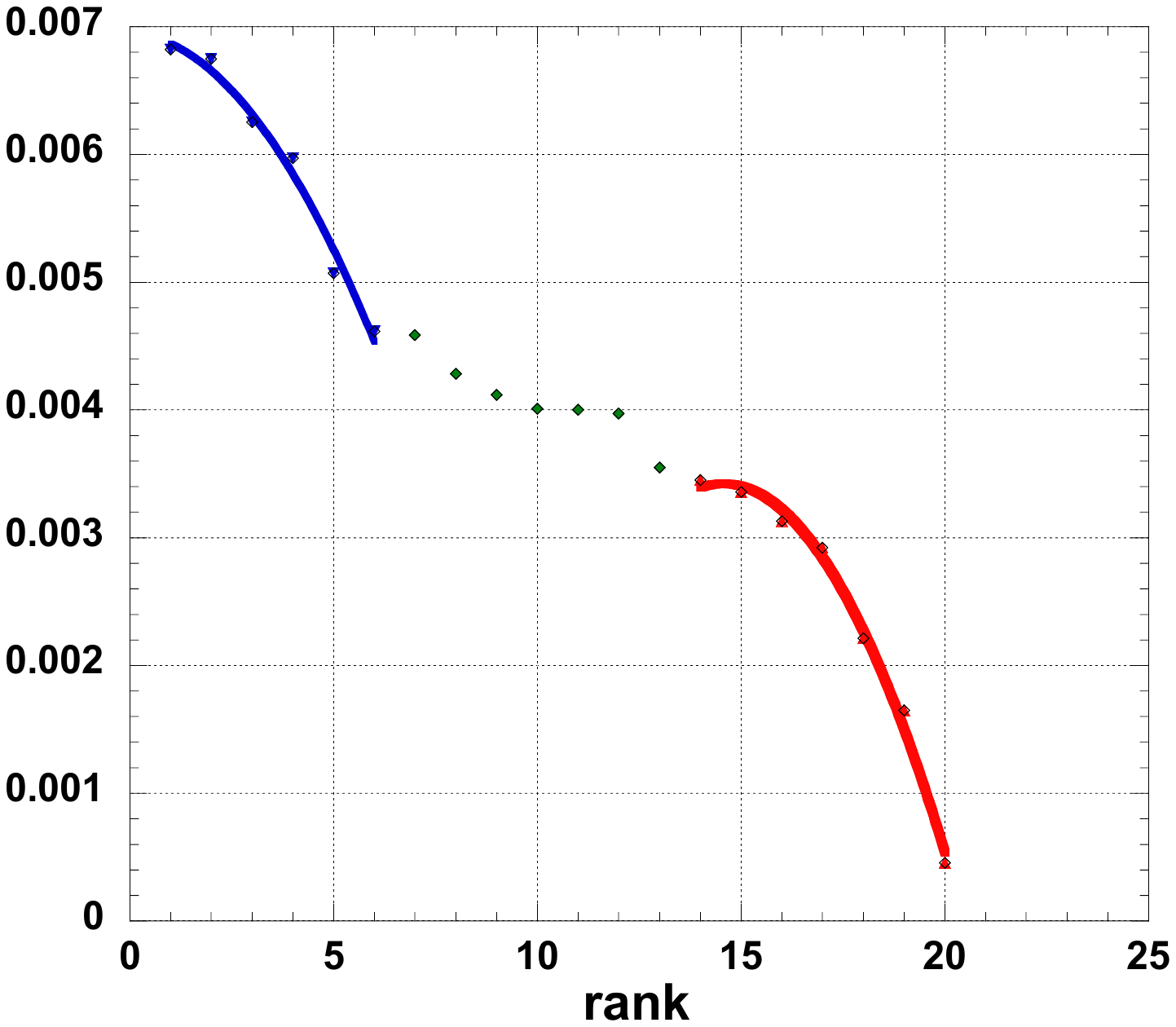}  % [height=14cm,width=15cm]
 \caption{Rank distribution of the "report distance  to disorder ", $\hat{d_j}$,  for the 20 reports, showing a bimodal distribution.  } \label{Plot3distanceplot} 
\end{figure}

Fig. \ref{Plot3distanceplot} shows the ranking of the distances, as a function of their rank.  The  distribution of  "distances" is seen to be   far from trivial. In particular, a shoulder is observed between two well marked bumps.   This is indicating  an interesting output of the investigation .   Therefore, a   distinction can again be made between two sets of reports, as  shown by a  simple fit (a quadratic law) to the low rank and high rank cases, respectively. It may be debated, in further work, whether  the intermediate set is  in fact to be considered as a  third set, or results from the other two.

%{\it I don't quite know if that helps, and if someone sees something, but the list of ranks is: $R_5$, $R_{18}$, $R_{10}$, $R_{14}$, $R_{16}$, $R_{15}$, $R_{13}$, $R_{12}$, $R_{19}$, $R_7$, $R_2$, $R_3$, $R_9$, $R_1$, $R_{20}$, $R_{11}$, $R_8$, $R_6$, $R_{17}$, $R_4$; thus $R_5$ is the most  "disordered" report; it has been shown above  to be the worse  report;   $R_4$ the less  disordered (most average !!!) report, and does not have many different words. This is also seen on other tables. }

\subsection{Report ranking distance  to journal  editorial standard }\label{ReportrankingSigma}

Another measure  can be proposed from

  \begin{equation} \label{Hinfinity}
  H ^{(\Sigma)}\equiv   -p(\alpha^{(\Sigma)})\; ln (p(\alpha^{(\Sigma)})),
  \end{equation}  
    where the value  $\alpha^{(\Sigma)}$ corresponds to the exponent relevant to the  whole  set of  reports, thus to a virtual reviewer, in some sense  characterizing the reviewers tied to the journal by the editors.  %Here, such an exponent  corresponds to  the case $R_{all}$, i.e.  $\alpha^{(\Sigma)}$=0.85. Since $\Sigma_j^{20} \alpha_j$= 0.812+0.841=1.653,  whence $  H^{(\Sigma)}=0.342$.   
     Similarly to Eq. (\ref{daverage}), let 
       %  $\alpha^{(\Sigma)}$=0.812, whence $  H^{(\Sigma)}=1.114 $.  or
  \begin{equation} \label{dinfinity}d_j^{(\Sigma)}=- \frac {H_j }{ln(N_j)}+ \frac {H^{(\Sigma)} }{ln(N_{\Sigma})}\end{equation}
   with $N\equiv N_{\Sigma}=1996$,  leading to  a set of $ d_i^{(\Sigma)} \simeq 0.123$. 
   
    The "final results"   for the 10 September and  for the 10 October reports are given in Table \ref{distancetableR1R10} and Table \ref{distancetableR11R20}  respectively. As could be expected, if the set of reviewers is rather homogeneous in behavior, %the  $d_j$  and  $d_j^{(\Sigma)}$  values are quite similar to each other: 
     the relative quantities follow a ranking similar to what is expected through the Zipf's law analysis. However, the   $\hat{d_j}$   and $ d_j^{(\Sigma)}$ values are not so coherent: the orders of magnitude differ, due to  a different  order of magnitude in the number of words, but also the different signs indicate different classes of reports (or reviewers). %Observe the  negative   $\hat{d_j}$  sign   and  the large negative $d_j^{(\Sigma)}$   value   for  $R_4$,  and for all  other September  $d_j^{(\Sigma)}$  cases.
%     \vskip0.4cm
  %   {\it I recommend to skip such Tables; they were interesting for me to see whether there was some "month effect" !  However, except for the last two columns, the other informations are already elsewhere. Moreover, the Tables can be misleading a reader, because the calculation of  "distances" depend on the set which is looked at; in fact, as I stressed elsewhere, these are not bona fide distances,  they are not additive, due to $ln$; they are rather  so called  "Kulback-Leibler divergence" measures; for completeness,one would have to show the equivalent of such tables, but when $p_{\alpha_j},$ $H_j,$  $d_j,$  $\hat{d_j}$ , $d_j^{(\Sigma)}$ values would result if/when  the 20 reports are seen as one group only. Thus I would simply stick to one Figure, likely Fig. \ref{Plot8djSigma} }

%{\bf MA comment : 
{\it In fine}, we stress that the interpretation of these two distances is different:     $\hat{d_j}$  measures a distance between reports, but  $ d_j^{(\Sigma)}$  pertains to a "more general" measure, with respect to a   virtual (average) reviewer  characterizing this  journal. From the latter, different journals could be compared.

	\begin{table} \begin{center}
     \begin{tabular}{|c|c|c|c|c|c|c|c|c|}
  \hline								
	$R_j$	& $\alpha_j$ & $ln(N_j)$	& $N_j$ & $p_{\alpha_j}$	& $H_j$ &	$d_j$ & $100 \hat{d_j}$ & 100$d_j^{(\Sigma)}$ \\ \hline			
$R_1$ & 0.708& 5.182 & 178 & 0.108& 0.241 & 0.954 & -0.073 & -0.145	\\
$R_2$ & 0.781 & 5.635 & 280 & 0.119 & 0.254 & 0.955 & 0.069 & -0.003	\\
$R_3$ & 0.635& 4.905 & 135 & 0.097 & 0.226 & 0.954 & -0.045& -0.116	\\
$R_4$ & 0.703& 4.673 & 107 & 0.107 & 0.240 & 0.949 & -0.557& -0.628	\\
$R_5$ & 0.622 & 5.371 & 215 & 0.095 & 0.224 & 0.958 & 0.407& 0.336	\\
$R_6$ & 0.714& 4.997 & 148 & 0.109 & 0.242 & 0.952 & -0.267 & -0.338	\\
$R_7$ & 0.595& 4.700 & 110 & 0.091& 0.218 & 0.954 & -0.070& -0.141	\\
$R_8$ & 0.651& 4.812 & 123 & 0.100 & 0.230& 0.952 & -0.202& -0.273	\\
$R_9$ & 0.564& 4.454 & 86 & 0.086 & 0.211 & 0.953 & -0.170 & -0.242	\\
$R_{10}$ & 0.567& 4.935 & 139 & 0.087& 0.212 & 0.957 & 0.275& 0.204	\\
\hline
\end{tabular}
\caption{The 10  Sept. reports ($R_i$) by reviewers  with their number of different words  $N_i$  and $ln(N_i)$ with  distance  measures, according to Sect. \ref{sec:entropy}..} \label{distancetableR1R10}
\end{center} \end{table}

	\begin{table} \begin{center}
     \begin{tabular}{|c|c|c|c|c|c|c|c|c|}
  \hline								
$R_j$	 & $\alpha_j$ & $ln(N_j)$ & $N_j$ & $p_{\alpha_j}$ & $H_j$ &	$d_j$ & 100$\hat{d_j}$ & 100$d_j^{(\Sigma)}$ \\ \hline				
$R_{11}$ & 0.841& 5.737& 310 & 0.112 & 0.245& 0.957& -0.137& 0.234	\\
$R_{12}$ & 0.767& 5.624& 277 & 0.102 & 0.233& 0.959 & -0.007& 0.365	\\
$R_{17}$ & 0.863&5.553& 258 & 0.115 & 0.248 & 0.955& -0.341 & 0.031\\
$R_{18}$ & 0.651& 5.526& 251 & 0.086 & 0.212& 0.962& 0.300& 0.671	\\
$R_{16}$ & 0.723& 5.565& 261 & 0.096& 0.225 & 0.960& 0.087 & 0.458	\\
$R_{13}$ & 0.750& 5.602& 271 & 0.099& 0.230 & 0.959& 0.029 & 0.401	\\
$R_{14}$ & 0.689& 5.572& 263 & 0.091& 0.219 & 0.961& 0.203& 0.575	\\
$R_{19}$ & 0.737& 5.447& 232 & 0.098& 0.227 & 0.958 & -0.046 & 0.325	\\
$R_{15}$ & 0.742& 5.568& 262 & 0.098& 0.228 & 0.959 & 0.030& 0.401	\\
$R_{20}$ & 0.770& 5.460& 235 & 0.102 & 0.233 & 0.957 & -0.141& 0.231	\\

\hline							
%$R_{11-20}$ & 0.841 & 7.3199 & 1510 & 0.11164 & 0.2448 & 0.9666 & -0.2241 & 0.1804	\\ \hline		
\end{tabular}
\caption{The 10  Oct. reports ($R_i$) by reviewers  with their number of different words  $N_i$  and $ln(N_i)$ with  distance  measures, according to Sect. \ref{sec:entropy}.} \label{distancetableR11R20}
\end{center} \end{table}

	\begin{table} \begin{center}
     \begin{tabular}{|c|c|c|c|c|c|c|c|c|}
  \hline								
	$R_j$	& $\alpha_j$ & $ln(N_j)$	& $N_j$ & $p_{\alpha_j}$	& $H_j$ &	$d_j$ &100$\hat{d_j}$ & 100$d_j^{(\Sigma)}$ \\ \hline														
$R_{1-10}$ & 0.812& 6.843 & 937 & 0.491  & 0.349 & 0.949 & -0.230 & -0.603	\\
$R_{11-20}$ & 0.841 & 7.320 & 1510 & 0.509 & 0.344& 0.953 & 0.177  & -0.196	\\
$R_{1-20}$ & 0.850 & 7.599& 1996	&	0.514 & 0.342 &0.955	& 0.373 &0
 	\\ \hline	
\end{tabular}
\caption{The 10  Sept., the 10 Oct.,  and all (10+10) Sept.+Oct. reports ($R_i$) by virtual reviewers  with their number of different words  $N_i$  and $ln(N_i)$ with  distance  measures, according to Sect. \ref{sec:entropy}.} \label{distancetableR1R20}
\end{center} \end{table}
%\clearpage

 \begin{figure}
%\centering
\includegraphics[height=14.8cm,width=14.8cm]{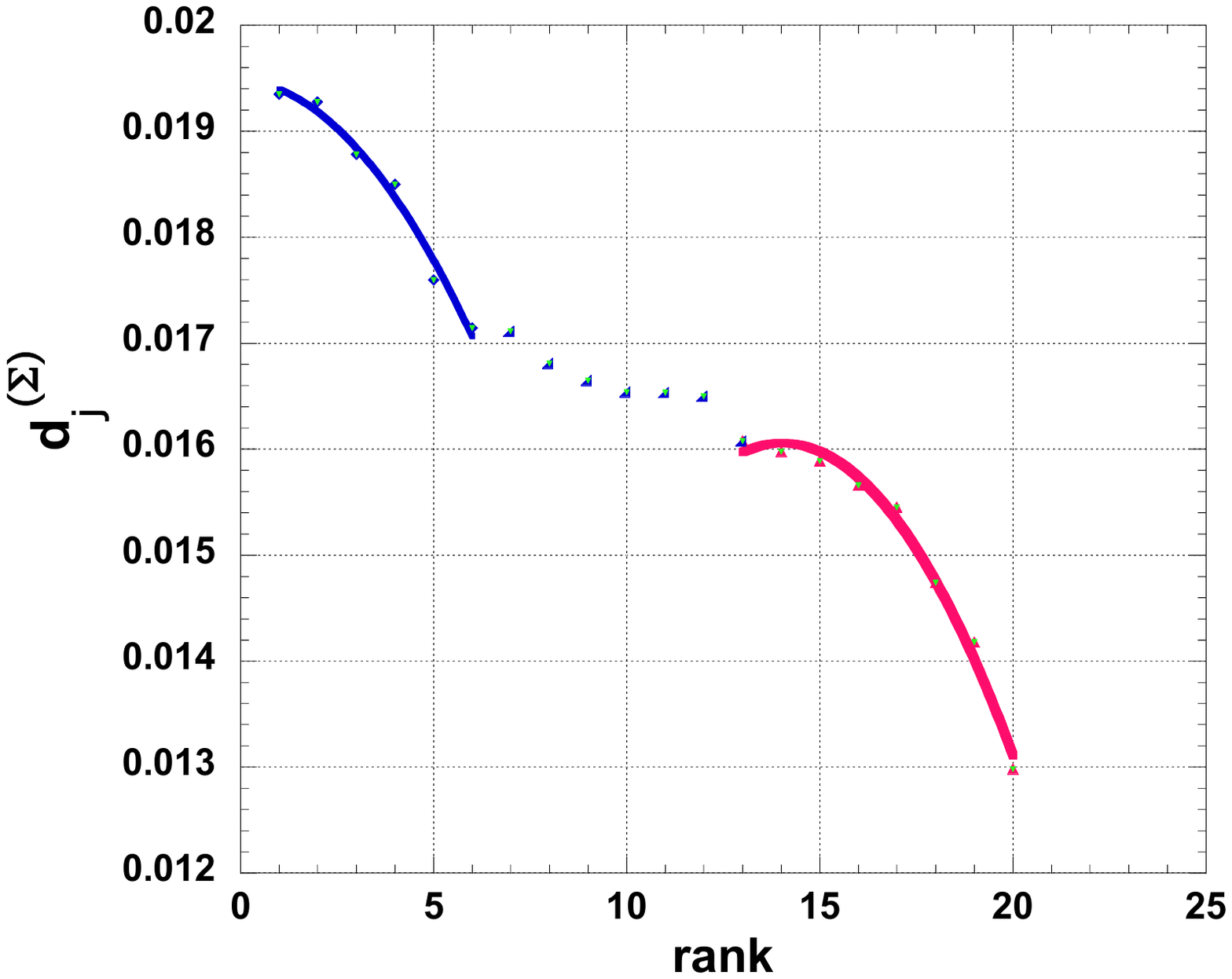}  % [height=14cm,width=15cm]
 \caption{Rank distribution of the  "distance  $d_j^{\Sigma}$  to disorder"  for the 20 reports, showing a bimodal distribution.   } \label{Plot8djSigma} 
\end{figure}
 
\section{Conclusions}\label{conclusions}

 This conclusion section, beside summarizing our findings and their possible limits, allows us to offer   a few suggestions for further research lines. To the best of our knowledge, this  seems to be the first time that  one quantifies reviewers through their report linguistic content.  Next, a few findings  are likely to be robust, others seem to be reliable.  This is encouraging because it suggests the feasibility of using simple quantitative measures to characterize various aspects related to the quality of reviews and reviewers. This opens the door toward developing complementary tools of automatic evaluations in parallel to peer review.
 
In  summary, this paper provides a statistical analysis of 25 reviewer reports for the biochemistry  section of respected chemistry  journal  JSCS  (e.g., impact factor= 0.912 in 2012). Each  report has been analyzed with respect to its word content, along the first Zipf's law idea  (Zipf 1949).
\begin{itemize} \item
It has been shown that the total number of words (TW) and of different words (DW)  depend entirely on reviewers, on their vocabulary, whence likely on their willingness to spend time on the peer-review process.
\item
 It is proven that  a  power law is appealing in describing the size-rank relationship. However, due to the value range of the Zipf exponent, such reports are found to be very different from usual texts by novelists  (Darooneh  and Shariati, 2014). 
 \item From the   Zipf exponent range, it seems that there are two classes of reports.
 \item It has been observed that the Zipf exponent  seems to characterize a specific reviewer, - though the statistics  could be  refined.
 \item  It is argued that one can compare reports (and maybe reviewers) between themselves through the concept of distance to randomness, basing our analysis on the thermodynamic entropy context, equivalent to the Shannon information entropy. 
  \item  It is argued that one can qualify reports (and maybe reviewers) with respect to journal standards, due to the choice of reviewers, through a similar concept of distance to randomness.
 \end{itemize}
 
  %\vskip0.5cm   
  Can we finally decide  whether the Zipf exponent characterizes the reviewer more than his/her  specific report ?  Aware  of  possibly to be  raised "ambiguities"  after our findings, it  can be considered that  many questions, thus suggestions for further work, follow. It can be  recommended that

 \begin{itemize}
  \item  more discussion can involve the acceptance/rejection effect  of papers; one approach could be  based on  correlations through quantified "linguistic aspects";  it is of common knowledge that rejected papers can contain quite sharp language
  \item    the statistics can be improved. %Even though the Zipf exponent  seems to characterize a specific reviewer, but. 
   However this  demands  much work, since every report must  be  reviewed for technical purposes as shown in the main text;
   
%   \item   difficult to know when the papers were rejected by one reviewer, if it is because the papers were not good enough or because of  the intrinsic nature of the reviewer
  \item the largest exponents might be correlated to the length  (or the number of words) of the report.
 \end{itemize}

 Several other  lines for further investigations  can be imagined to arise  (mainly) from restrictions at the start of the data acquisition process. It is suggested that 
 \begin{itemize}
 \item  more samples,  - in order to reduce the error bars
  \item   other sub fields of chemistry and in other scientific fields,   - in order to test some universality, if any  
\item   other journals,  - in order to test some universality, if any  
 \item  other quantifying techniques,
 \end{itemize}
   could  be examined.

   Last but not least, an approach on the peer review dynamics is of interest, i.e., on the  willingness  of reviewers to spend time on such peer-review reports, when asked.  Finding a hierarchy of reviewers  within the present aim of paying  reviewers is certainly worthwhile to be examined.
 
% olgica :  Reviewers' comments are in word file (that's how we receive them), and they are listed as a collection, differentiating between papers which have editorial numbers, reviewers for one manuscript and, as Piotr asked, I graded reports according to what authors have gained out of them. As expected, longer reports were more detailed and more beneficial to authors. And yes, I also consider JSCS data as model system. I do not, actually, believe, that much different results would be obtained by examining reports from other journals
  \vskip0.5cm
 {\bf Acknowledgements}
 This paper is part of scientific activities in COST Action  TD1306
New Frontiers of Peer Review (PEERE).

\vskip0.5cm %\newpage
{\bf Appendix A.  Data analysis of unmanipulated reports}
\vskip0.5cm
In this Appendix, the  data analysis of the 10  Sept.  reports,  without any data manipulation, i.e. without  in any way modifying  the reports for their word content,  is reported. It is  shown in Table \ref{rawdatapwlfitparam} that the  power law exponent  appears to be  $\simeq 0.781\pm 0.004$, with regression   coefficient $R^2 \in 0.989$, for the  $R_2$ report.

For $R_4$,   $R_5$, and  $R_6$, the  power law exponent  appears to be  $\simeq 0.70\pm 0.01$, with their regression   coefficient $R^2 \in 0.966, 0.938$,
 but  the other exponents evenly spread on 0.423 till 0.683.

It  was observed that  $R_2$, $R_5$ and $R_6$ are  among the top  three longest reports. % ($R_3$ is the third one). 
Thus should be less sensitive to "slight" data modifications.

In fact, this Table,  through the found  $\alpha$ values,  is proving that for short reports one has to be  much concerned by the vocabulary; whence one has to distinguish the meaning of  (short) words (like "a", "c", "k").

%By the way, the Zipf analysis of the raw data for the "sum" of the 10 reports gave  an exponent  0.8515, with a regression coefficient $R^2$= 0.9815.

 In conclusion,  it is highly meaningful to "adapt" (= rewrite) the reports. The bad thing is that life is not simplified from a scientific point of view since it takes time to rewrite reports in a useful way.

\begin{table}[htdp]
\begin{center}
\begin{tabular}{|c|c|c|c|c|c|c|c|c|}
\hline
                     &                      $S_1$                 &                  $\alpha$                  	&     $\chi^2$	&     $R^2$	\\  \hline
$R_9$	&	 8.74$\pm$0.27&	0.556$\pm$0.014  &	10.31	&	0.922	\\
$R_4$	&	18.39$\pm$0.35&	0.704$\pm$0.011 &	18.93	&	0.966 	\\
$R_7$	&	11.73$\pm$0.23&	0.604$\pm$0.009 &	9.168	&	0.962 	\\
$R_8$	&	37.49$\pm$1.01&	0.622$\pm$0.011 &	358.1	&	0.902	\\
$R_3$	&	16.71$\pm$0.23&	0.635$\pm$0.007 &	11.34	&	0.978 	\\
$R_{10}$	&	15.86$\pm$0.38&	0.567$\pm$0.010 &	33.19	&	0.940 	 \\
$R_6$	&	35.02$\pm$0.63&	0.712$\pm$0.010 &	84.12	&	0.964 	\\
$R_1$	&	30.27$\pm$0.31&	0.708$\pm$0.005 &	24.91	&	0.986	\\
$R_5$	&	12.57$\pm$0.30&	0.651$\pm$0.012 &	16.53	&	0.938	\\
$R_2$	&	63.78$\pm$0.46&	0.781$\pm$0.004 &	80.57	&	0.989 \\  \hline
$R_{all}$	&	242.45$\pm$2.05&	0.806$\pm$0.005 &	1683.3	&	0.985	\\  \hline
\end{tabular}
\caption{  Power law fit parameter, Eq.(\ref{Zipfeq}),  for the 10  Sept. "unmanipulated" reports  $R_i$.} 
\label{rawdatapwlfitparam}
\end{center} \end{table}

\vskip0.5cm %\newpage
%\begin{thebibliography}{99}
{\bf References}
\\Ê\\ 
  Ausloos M.  2012a   Generalized Hurst exponent and multifractal function of original and translated texts mapped into frequency and length time series. {\it Phys. Rev.  E}  {\it 86}, 031108. 
 \\Ê\\
 Ausloos M.  2012b    Measuring complexity with multifractals in texts. Translation effects, {\it Chaos, Solitons and Fractals} {\it 45},  1349-1357.  
\\Ê\\ %\bibitem{Ausloos_2013} 
Ausloos M.    2013   A scientometrics law about co-authors and their ranking. The co-author core, {\it  Scientometrics}   {\it  95},   895-909.
\\Ê\\ %\bibitem{Bornmann_2011} 
 Bornmann L.,    2011   Scientific peer review,  {\it  Ann. Rev. Inf. Sci. Technol.},  {\it 45},   199-245.
\\Ê\\ %\bibitem{Bougrine_2014} 
  Bougrine H.,    2014   Subfield Effects on the Core of Coauthors, {\it  Scientometrics},   {\it 98},   1047-1064.
\\Ê\\ %\bibitem{Callaham_1998} 
 Callaham M.L., Wears R.L., Waeckerle J.F.,    1998   Effect of attendance at a training session on peer reviewer quality and performance, {\it  Ann. Emerg. Med.},   {\it 32},   318-322.
 	\\Ê\\ %\bibitem{Pietronero_2012} 
 Cristelli M., Batty M., Pietronero L.,   2012   There is more than a power law in Zipf, {\it  Scientific Reports},  {\it 2}, 812.	doi:10.1038/srep00812
 \\Ê\\
 Darooneh A.H.,  Shariati, A.,  2014   Metrics for evaluation of the author's writing styles: Who is the best?, {\it  Chaos: An Interdisciplinary Journal of Nonlinear Science},   {\it 24},   033132.
\\  \\ Dubois D.M.,  2014   Computational Language Related to Recursion, Incursion and Fractal in {\it  Language and Recusrsion} F. Lowenthal and  L. Lefebvre  eds.) Springer Science+Business Media, New York, pp. 149-165.
 \\Ê\\ %\bibitem{Fairthorne_1969} 
 Fairthorne R.A.,  1969   Empirical hyperbolic distributions  Bradford-Zipf-Mandelbrot) for bibliometric description and prediction, {\it  J. Documentation},   {\it 25},   319-343.
  \\Ê\\
 Febres  G., Jaffe K.  2014   Quantifying literature quality using complexity criteria,
 preprint $arXiv:1401.7077$.
  \\Ê\\ÊFerrer i Cancho R,   2006   When language breaks into pieces. A conflict between communication through isolated signals and language. {\it Bio Systems}, {\it 84}, 242--253.
\\Ê\\ %\bibitem{Feurer_1994} 
Feurer I.D., Becker G.J., Picus D., Ramirez E., Darcy M.D., Hicks M.E.,   1994   Evaluating peer reviews: pilot testing of a grading instrument,  {\it  JAMA}   {\it 272},   98-100.
\\Ê\\ %\bibitem{Godlee_1998} 
Godlee F., Gale C.R., Martyn C.N.,    1998   Effect on the quality of peer review of blinding peer reviewers and asking them to sign their reports: a randomized control trial.  {\it  JAMA}   {\it 280},   237-240.
\\Ê\\ %\bibitem{Goodman_1994} 
Goodman S.N., Berlin J., Fletcher S.W., Fletcher R.H.,    1994   Manuscript quality before and after peer review and editing at Annals of Internal Medicine,  {\it  Ann. Intern. Med.}   {\it 121},   11-21. 
\\Ê\\ %\bibitem{Hill_2004} 
 Hill B.M.,   2004   The Rank-Frequency Form of Zipf's Law,  {\it  J. Am. Stat. Assoc.},   {\it 9},   1017-1026. 
\\Ê\\ %\bibitem{Jadad_1998} 
Jadad A.R., Cook D.J., Jones A., Klassen T.P., Tugwell P., Moher M., Moher D.,    1998   Methodology and reports of systematic reviews and meta-analyses: a comparison of Cochrane reviews with articles published in paper-based journals. {\it   JAMA},    {\it 280},   278-280.
\\Ê\\ %\bibitem{Justice_1998}  
Justice A.C., Cho M.K., Winker M.A., Berlin J.A., Rennie D.,    1998   Does masking author identity improve peer review quality? a randomized controlled trial. {\it   JAMA},    {\it 280},   240-242.
\\Ê\\ %\bibitem{Laherrere_1998}
 Laherrere J., Sornette D.,     1998   Stretched exponential distributions in nature and economy fat tails with characteristic scales. {\it   Eur. Phys. J. B},    {\it 2},   525-539.
\\Ê\\ %\bibitem{Lin_2010}
 Lin S.,   2010   Rank aggregation methods.  {\it   WIREs Comp. Stat.},    {\it 2},   555-570.
\\Ê\\ %\bibitem{McCowan_2002} 
McCowan B., Doyle L.R., Hanser S.F.,     2002   Using Information Theory to Assess the Diversity, Complexity, and Development of Communicative Repertoires. {\it  J. Comp. Psychol. } {\it 116}, 166-.
\\Ê\\ %\bibitem{McKean_2009} 
McKean J.W., Terpstra J.T., Kloke J.D.,  2009   Computational rank-based statistics. {\it   WIREs Comp. Stat.},    {\it 2},   132-140.
\\Ê\\ %\bibitem{McNutt_1990} 
McNutt R.A., Evans A.T., Fletcher R.H., Fletcher S.W.,     1990   The effects of blinding on the quality of peer-review. a randomized trial. {\it   JAMA},    {\it 263},   1371-1376.
\\Ê\\ %\bibitem{Miskiewicz_2013} 
 Miskiewicz J.,   2013    Effects of Publications in Proceedings  on the Measure of the Core Size of Coauthors. {\it  Physica A},    {\it 392},   5119-5131.
\\Ê\\ %\bibitem{Neuhauser_1989} 
Neuhauser D., Koran C.J.,   1989   Calling Medical Care reviewers first: a randomized trial. {\it  Med. Care},    {\it 27},   664-666.
\\Ê\\ %\bibitem{Oxman_1991} 
Oxman A.D., Guyatt G.H., Singer J.,   1991   Agreement among reviewers of review articles. {\it   J. Clin. Epidemiol.},   {\it 44},  91-98.
\\Ê\\ %\bibitem{PRC_2008} 
Publishing Research Consortium,    2008    {\it  Peer review in scholarly journals: Perspective of the scholarly community: an international study}.
\\Ê\\Ê
Rodriguez E.,   Aguilar-Cornejo M.,   Femat R.,   Alvarez-Ramirez J.,  2014   Scale and time dependence of serial correlations in word-length time series of written texts. {\it Physica A} {\it  414},  378--386.
\\Ê\\ %\bibitem{Rooyen_1999} 
van Rooyen S., Godlee F., Evans S., Black N., Smith R.,    1999   Effect of open peer review on quality of reviews and on reviewers' recommendations: a randomised trial. {\it   BMJ},    {\it 318},   23-27.
\\Ê\\ %\bibitem{shannon48} 
  Shannon C.,    1948   A mathematical theory of communications.  {\it  Bell System Technical Journal},    {\it 27},  379-423.
\\Ê\\ %\bibitem{shannon51} 
 Shannon C.,  1951   Prediction and entropy of printed English.  {\it  Bell System Technical Journal},   {\it 30}, 50-64.	
\\Ê\\ %\bibitem{Siler_2015} 
Siler K., Lee K., Bero L.,   2015    Measuring the effectiveness of scientific gatekeeping. {\it   Proc. Nat. Acad. Sci.},   {\it 112},  360-365.
\\Ê\\ %\bibitem{Strayhorn_1993}
 Strayhorn Jr J., McDermott Jr J.F., Tanguay P.,   2015     An intervention to improve the reliability of manuscript reviews for the Journal of the American Academy of Child and Adolescent Psychiatry. {\it  Am. J. Psychiatry},    {\it 150},  947-952.
\\Ê\\ %\bibitem{Wager_2001} 
 Wager E., Jefferson T.,  2001   The shortcomings of peer review. {\it   Learned Publishing},    {\it 14},   257-263.
\\Ê\\ %\bibitem{Wieder_2009}
 Wieder T.,    2009   The Number of Certain Rankings and Hierarchies Formed from Labeled or Unlabeled. {\it   Appl. Math. Sci.},   {\it 3}, 2707-2724.
\\Ê\\ %\bibitem{Wolfe_2009} 
Wolfe D.A.,  2009   Rank methods. {\it    WIREs Comp. Stat.},   {\it 2},   342-347.
\\Ê\\ %\bibitem{Wolfe_2010} 
Wolfe D.A.,  2010   Ranked set sampling. {\it     WIREs Comp. Stat.},    {\it 2},   460-466.
\\Ê\\ %\bibitem{Zipf_1949} 
Zipf G.K.,    1949    {\it  Human Behavior and the Principle of Least Effort : An Introduction to Human Ecology}, Cambridge, Mass.: Addison Wesley Press.

%\end{thebibliography}

\end{document}